\documentclass{iopart}

\usepackage[dvips]{graphicx}
\usepackage{amssymb}
\usepackage{color}
\usepackage{bm}

\eqnobysec

\def\n*{n_{*}}
\def\tm{\tilde{\mu}}

\begin{document}

\title{Beliaev technique for a weakly interacting Bose gas } 


\author{B~Capogrosso-Sansone,$^{1,2}$ S~Giorgini,$^3$ S~Pilati,$^{3,4}$ L~Pollet,$^{1,5}$ N~Prokof'ev,$^{1,6}$ B~Svistunov$^{1,4,6}$ and M~Troyer$^4$}

\address{$^1$ Department of Physics, University of Massachusetts, Amherst, MA 01003, USA}

\address{$^2$ Institute for Theoretical Atomic, Molecular and Optical Physics, Harvard-Smithsonian Center of Astrophysics, Cambridge, MA, 02138, USA}

\address{$^3$ Dipartimento di Fisica, Universit\`a di Trento and CNR-INFM BEC Center, I-38050 Povo, Trento, Italy}

\address{$^4$ Theoretische Physik, ETH Zurich, CH-8093 Zurich, Switzerland}

\address{$^5$ Physics Department, Harvard University, Cambridge, MA 02138, USA}

\address{$^6$ Russian Research Center ``Kurchatov Institute'', 123182 Moscow, Russia}

 \ead{svistunov@physics.umass.edu}


\begin{abstract}
Aiming for simplicity of explicit equations and at the same time controllable accuracy of the theory we present results for all thermodynamic quantities and
correlation functions for the weakly interacting Bose gas at short-to-intermediate distances obtained within an improved version of Beliaev's diagrammatic technique.
With a small symmetry breaking term Beliaev's diagrammatic technique becomes regular in the infrared limit. Up to higher-order terms (for which we present order-of-magnitude estimates), the partition function and entropy of the system formally correspond to those of a non-interacting bosonic (pseudo-)Hamiltonian with a temperature dependent Bogoliubov-type dispersion relation. Away from the fluctuation region, this  approach provides the most accurate---in fact, the best possible within the Bogoliubov-type pseudo-Hamiltonian framework---description of the system with controlled accuracy. 
It produces accurate answers for 
the off-diagonal correlation functions up to distances where the behaviour of correlators is controlled by generic hydrodynamic relations, and thus can be accurately extrapolated to arbitrarily large scales. 
In the fluctuation region, the non-perturbative contributions are given by universal (for all weakly interacting U(1) systems) constants and scaling functions, which can be obtained separately---by  simulating classical U(1) models---and then used to extend the description of the weakly interacting Bose gas to the fluctuation region.
The theory works in all spatial dimensions and we explicitly check its validity against first-principle Monte Carlo simulations for various thermodynamic properties and the single-particle density matrix.
\end{abstract}

  \maketitle

\section{Introduction}
\label{sec:I}

For nearly half a century the theory of the weakly interacting Bose gases (WIBG)
remained in the realm of purely theoretical investigations
\cite{Bogoliubov,Beliaev,Pines,Hohenberg65,Popov87,Kagan87,Fisher}
(for a recent review, see \cite{Andersen})
providing insight into the nature of superfluid states of matter but not directly
relating to existing experimental systems. The situation changed
with the realization of Bose Einstein condensation in cold atomic gases
\cite{Cornell,Ketterle,Bradley}. The typical values of the interaction
parameter for alkali atoms are very small $na^3\sim 10^{-6}$,
where $n$ is the number density of the gas and $a$ is the $s$-wave scattering length.
Since 1995 both the theory of WIBG and experiments have progressed
with a close relationship between the two~\cite{PS03}.

Until recently, existing mean-field and variational
treatments such as the Bogoliubov zero-temperature approximation and the
finite-temperature quasi-condensate theory \cite{Svistunov}
were capable of describing the data within the relatively large experimental
uncertainties. However, the improvements in detection techniques, the studies of optical lattice systems
in which the effective gas parameter can be made arbitrarily large,
and the need for reliable thermometry, e.g., through precise entropy matching,
all indicate the need for a more accurate and controllable theory. It is highly desirable, similarly to
the well-known corrections to the ground state energy \cite{Yang57} and
condensate density \cite{Bogoliubov}, to have a theory which accounts
for leading corrections to all thermodynamic properties at finite temperature, works in any dimension, and
provides estimates for omitted higher order terms.

The accuracy of the description plays an important role in this paper. We systematically estimate the errors of all approximations, which is
crucial for comparing seemingly different schemes proposed for Bose gases.
Indeed, alternative theories can be equivalent to each other within the level of their accuracy,
but a definitive conclusion can only be drawn on the basis of analyzing their
systematic errors. For example, within the same accuracy gap-less and gap-full finite-temperature
schemes may often be regarded as equivalent without any preference for one of them, as long as
the gap is on the order of or smaller than the terms in the chemical potential which the theory ignores.
Also, in the fluctuation region all perturbative schemes are equally
inaccurate in terms of their treatment of  the order parameter, and it makes no sense
whatsoever to distinguish between them by the type of the transition they predict.

In this work, we provide a rigorous framework for obtaining a consistent description of all thermodynamic finite-$T$ properties of the WIBG in one, two, and three dimensions.
It is based on Beliaev's regularized diagrammatic technique \cite{Beliaev} which allows us to
calculate all relevant thermodynamic functions of the system. The diagrammatic technique also leads to an accurate description of correlation functions up to sufficiently large distances where Popov's hydrodynamic description takes over \cite{Popov87}. The key results of our calculations are summarized in section \ref{sec:summary}, which is accessible to the general reader.

It turns out that all final results for thermodynamic functions perfectly fit into the pseudo-Hamiltonian
picture,  i.e. we prove that the same answers would be obtained if they were calculated
in the standard way with Bogoliubov's non-interacting quasi-particle picture (with a self-consistently defined temperature dependent spectrum).

In addition, we find an explicit expression for the pressure and give a more accurate expression for the chemical potential. We obtain the equations of state for all basic thermodynamic quantities in an integral parametric form using temperature, effective chemical potential, and the interaction strength as parameters.
These results include sub-leading corrections \footnote{The only exceptions are the quantities, like entropy and heat capacity, that vanish as $T\to 0$. At low enough temperatures, 
these quantities lose their ideal-gas leading terms,
getting nothing instead---as opposed to energy, chemical potential, and pressure,  having straight-forward mean-field contributions that start to lead at low temperatures.}. The analysis of systematic errors shows that the accuracy of our results cannot be further improved within the paradigm of non-interacting quasiparticles with a (temperature-dependent) Bogoliubov spectrum. Moreover, there exists a two-parametric continuum of possibilities for recasting the form of the final answers with the same accuracy due to the freedom of modifying the definitions of  approximate notions of ``effective interaction'' and ``effective chemical potential''. Using this freedom, one can trade the higher order terms in the explicit integral expressions for the thermodynamic quantities for those of the effective interaction vertex. Our analysis shows that the question of the energy and momentum dependence of the interaction vertex is more a matter of taste rather than accuracy. We show that even in 2D---where, in view of the vanishing $s$-scattering amplitude, a low-energy cutoff is unavoidable and one could expect that the momentum-dependent effective-interaction vertex is crucially important in any theory taking care of sub-leading corrections---the effects associated with the energy and momentum dependence of the effective interaction can be naturally accounted for on equal footing with all the other effects of the same order. Within the systematic error bars of our pseudo-Hamiltonian approach, the effective interaction of the 2D WIBG can be safely treated as an energy and momentum independent constant.

\begin{figure}
\includegraphics[angle=-90, scale=0.3, bb=0 -250 600 500]{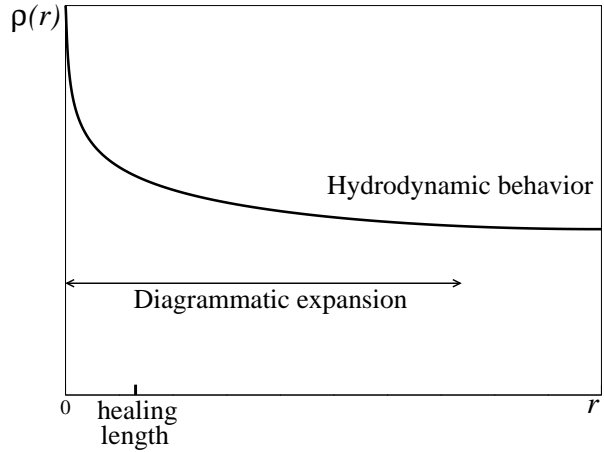}
\caption{ Sketch of the density matrix dependence on distance in the WIBG at temperatures below
the fluctuation region. After an initial drop the density matrix
levels-off at the healing length and undergoes slow decay up to a distance where
the superfluid hydrodynamic approach becomes applicable. The U(1) symmetry breaking terms
make the density matrix to converge to a finite value (in any dimension)
at the hydrodynamic length scales.}
\label{fig:bimodal}
\end{figure}

Such analytical approaches are not supposed to work in the vicinity of the critical point $T_{\rm c}$. This so-called fluctuation region is characterized by
scaling functions that are universal for all U(1) weakly-interacting systems. It has already been studied numerically with high accuracy \cite{Tc3D,Arnold,Tc2D}, 
so that a combination of analytical and numerical data provides an accurate description at all temperatures.

Despite its success in treating the weakly interacting Bose gas, Beliaev's original technique features a subtlety that has seriously questioned its reliability: the normal and anomalous self energies---the fundamental quantities of the theory---have been demonstrated to be essentially momentum-dependent, with the anomalous self-energy vanishing in the limit of zero momentum and frequency \cite{Nepom78,Popov79,Nepom83}. On the other hand the first-order diagrams in the theory of the weakly interacting gas with a contact interaction yield momentum independent self energies. In the literature, one can find a number of recipes of how to deal with this inherent infrared problem of Beliaev's technique. In Popov's approach \cite{Popov87}, the
diagrammatic technique applies only to the higher-momentum part of the bosonic field, while the lower-momentum part is treated separately within the hydrodynamic representation.
Similar results can be achieved by working with a finite-size system and taking advantage of the bi-modality of correlation properties of weakly interacting superfluid bosonic systems (away from the fluctuation region) \cite{KSS}. The bi-modality allows one to select such a system size that in terms of local thermodynamic properties the system is already well in the macroscopic limit while the effect of the infra-red renormalization of the self energies is still negligibly small. Another option of a controllable microscopic description of weakly interacting Bose gas is the renormalization-group treatment \cite{Pistolesi,Dupuis07}. Similar to the finite-size regularization, our approach is to exploit the bi-modality
of correlation properties of the WIBG, see figure~\ref{fig:bimodal},  by cutting off infra-red divergencies using small U(1) symmetry breaking terms. The amplitude of the symmetry-breaking terms is such that their effect on the local thermodynamic properties of the system is negligibly small while all infrared singularities in Beliaev's technique are gone. Clearly, the  solution to the infrared problem comes at the expense of suppressing long-range fluctuations of the phase of the order parameter (with opening of a gap in the spectrum at small momenta), and thus distorting the long-range behaviour of correlation functions. Nevertheless, this distortion
is essentially irrelevant since it takes place at large distances where the (generic to all superfluids) behaviour of correlators is governed by hydrodynamic fluctuations of the phase  of the order parameter, with a simple and transparent theoretical description \cite{Giorgini}. Moreover, in all final answers one can explicitly take
the limit of vanishing symmetry breaking terms.

In terms of the final results, we confine our analysis to the most typical (and most difficult in two and three dimensions) case of the dilute gas, when the size of the potential $R_0$ is much smaller than 
the  distance between the particles:
\begin{equation}
 R_0 \ll n^{-1/d} .
\label{dilute}
\end{equation}
In principle, the theory works also at $R_0 \gtrsim n^{-1/d}$, and becomes
even simpler, since in this case the weakness of interaction literally implies Born type of the potential (while in the opposite case the amplitude of
potential can be arbitrarily large in two and three dimensions). However, only with the condition (\ref{dilute}) the final answers become insensitive
to the details of the interaction potential. The second condition which is assumed in our final results is
\begin{equation}
\lambda_T \gg R_0 , 
\label{lambda_T_vs_R_0}
\end{equation}
where $\lambda_T$ is the de Broglie wavelength.
Given the inequality  (\ref{dilute}), the condition (\ref{lambda_T_vs_R_0}) is  satisfied  {\it automatically} as long as one is interested 
in essentially quantum properties of the system, implying $\lambda_T^d n \gtrsim 1$. An extension of the theory to the Boltzmann 
high-temperature regime $\lambda_T \lesssim R_0$, where answers become sensitive to the particular form of the interaction potential, goes beyond the scope of this paper.

The paper is organized as follows. In section \ref{sec:summary} we first summarize the key results of our approach for thermodynamic quantities and correlation functions in a form easily accessible to the general reader. The next chapters systematically introduce the technique and derive the results. In section \ref{sec:Diagrams} we re-derive the diagrammatic
technique for the phase with the broken U(1) symmetry by introducing explicit (arbitrary small)
symmetry breaking terms in the Hamiltonian. We keep the discussion as general as possible to cover
inhomogeneous, e.g.\ trapped, systems and states with non-zero average current. In section \ref{sec:Diagrams} we also derive the central expressions for the proper self energies and chemical potential.
In section \ref{sec:Quasicondensate} we consider thermodynamics (including the superfluid density) of a homogeneous system in the superfluid phase. In particular, we introduce the notion of the renormalized chemical potential $\tilde{\mu}$ and show that $\tilde{\mu}$ and temperature form a natural pair of thermodynamic parameters, in terms of which all the other quantities are obtained from
the single-particle momentum-space integrals. In section \ref{sec:Estimates} we discuss, at the order-of-magnitude level, the structure of higher-order corrections to the spectrum of elementary excitations and thermodynamic quantities. This analysis, in particular, leads to the conclusion that within the pseudo-Hamiltonian approximation and Bogoliubov ansatz for the spectrum of elementary excitations our results cannot be further improved. In section \ref{sec:off} we show that the answers for the asymptotic long-range behaviour of off-diagonal correlation functions are readily obtained on top of Beliaev's diagrammatic treatment for medium-range correlations, by employing the generic hydrodynamic description of a superfluid.
The short section \ref{sec:Normal_region} deals with the normal region.
In section \ref{sec:Pseudo} we show that in both normal and superfluid regimes (but away from the fluctuation region around the critical point) the thermodynamic relations can be associated with a pseudo-Hamiltonian in the following sense. The partition function and the entropy of the system turn out to correspond to those of a non-interacting bosonic Hamiltonian with temperature-dependent parameters and a temperature-dependent global energy shift. In section \ref{sec:fluct} we render the results for the fluctuation region,
which previously were obtained in 2D and 3D, but not in 1D.
In section \ref{sec:Numerics} we finally compare our solutions with first-principle Monte Carlo data for WIBG both in continuous space and on a lattice.

\section{Key results}
\label{sec:summary}

Here we summarize the most important relations which we derive in the main part of the paper, assuming that the conditions (\ref{dilute}) and (\ref{lambda_T_vs_R_0})
are met.   The results of this section work for both continuous-space and lattice systems; in the latter case,
the mass $m$ is understood as the effective mass of low-energy motion. The equilibrium state of the system is 
conveniently parametrized by two independent variables, the temperature $T$ and the effective chemical potential $\tm$. The latter is always negative, so that $\tm \equiv - |\tm|$.

\subsection{Coupling constant}

The interaction is characterized by an (effective) coupling constant $U$.  In one dimension $U$ is the zero-momentum Fourier component of the bare potential ${\cal U}({\bf r})$. In two and three dimensions, $U$ is naturally expressed in terms of the $s$-wave scattering length $a$; the latter being defined as the radius of hard disk/sphere potential 
\begin{equation}
{\cal U}_{\rm hard}(r) \; = \;\left\{ {\begin{array}{*{20}c}
\infty,~~~~~~r<a\, ,  \\
~0,~~~~~~~r\geq a\, ,
\end{array}} \right.
\label{hard_pot}
\end{equation}
with low-energy scattering properties identical to those of the 
potential ${\cal U}({\bf r})$. A representational freedom allows one to use slightly different expressions for $U$, without sacrificing the order of accuracy. Our particular choice is
\begin{equation}
U  \, = \, {4\pi a\over m}~~~~~(d=3) \,  ,
\label{U_3D_gas}
\end{equation}
in three dimensions, and
\begin{eqnarray}
U = {4\pi/m \over \ln (2/a^2m\epsilon_0) -2\gamma}  ~~~~~(d=2) \,  ,
\label{U_2D_gas}
\end{eqnarray}
in two dimensions,
where $\gamma=0.5772\dots$ is Euler's constant. A particular expression for $U$ comes from fixing the specific form of an auxiliary function $\Pi(k)$.
The latter  enters thermodynamic relations in such a way that the values of those remain insensitive---up to irrelevant higher-order corrections---to 
slightly changing $U$ by adopting a slightly different $\Pi$. Equations (\ref{U_3D_gas}) and (\ref{U_2D_gas}) correspond to
\begin{equation}
\Pi (k)  \, = \, -{1\over 2 \, \epsilon(k)}~~~~~(d=3) \,  ,
\label{Pi_3D_case}
\end{equation}
\begin{equation}
\Pi (k)  \, = \, -{1\over 2} \, {1\over \epsilon(k) + \epsilon_0}~~~~~(d=2) \,  ,
\label{Pi_2D_case}
\end{equation}
respectively, where $\epsilon(k)$ is the particle dispersion relation. In one dimension $\Pi \equiv 0$. Within the representational freedom, different choices of the $(\Pi,U)$-pair 
are straightforwardly connected with  each other by (\ref{U1_U2})-(\ref{C_12}).

In the 2D case, the coupling constant $U$, and, correspondingly, function $\Pi(k),$ slowly evolve with changing effective chemical potential  and temperature: 
\begin{equation}
\epsilon_0\, \equiv\, \epsilon_0 (\tm,T) \; = \;\left\{ {\begin{array}{*{20}c}
({\rm e}/2)|\tm|~~~(T<T_{\rm c}),
 \\
~~~T~~~~~~~~~(T>T_{\rm c}) .
\end{array}} \right.
\label{epsilon_0_mu_T}
\end{equation}
It is {\it well} within the representational freedom to multiply the r.h.s. of (\ref{epsilon_0_mu_T}) by a constant of order unity, provided  the same expression for $\epsilon_0$
is used in both (\ref{U_2D_gas}) and (\ref{Pi_2D_case}). The only reason for introducing the order-unity factor ${\rm e}/2$ is to cast the zero-temperature equations of state in the form
(\ref{2D_ground_1})-(\ref{2D_ground_2})  adopted previously by some authors.

In the case when the problem of finding the scattering length $a$ is not straightforward, one can
resort directly to the integral equation (\ref{eq:def_of_U}) defining $U$ in terms of ${\cal U}$ and $\Pi$. Numeric solution of this equation can be obtained 
within bold diagrammatic Monte Carlo
technique \cite{boldMC}, which gives the scattering length as a by-product of the calculation.

\subsection{Thermodynamics in the superfluid region}
Thermodynamic quantities are obtained in parametric form, as functions of the pair of parameters $\tm$ and $T$.
The main relation is for the number density:
\begin{equation}
n(\tm, T)\, =\, |\tm|/U+ n' \, ,  
\label{number_density}
\end{equation}
\begin{equation}
n' = \sum_{\bf k} \left[  
{\epsilon(k) - E(k) \over 2E(k)} -|\tm|\Pi(k) +
{\epsilon (k) \over E(k)}\, N_E\right]  \, , \label{n_prime_rel}
\end{equation}
where
\begin{equation}
E(k)\, =\, \sqrt{\epsilon(k)[\epsilon(k)+ 2|\tm|] }
\label{Bog_disp}
\end{equation}
is the Bogoliubov-type  quasiparticle dispersion relation  and 
\begin{equation}
N_E \, =\, \left( {\rm e}^{E/T} -1 \right)^{-1} 
\label{Bose_Einstein}
\end{equation}
is the Bose-Einstein occupation number. In equation (\ref{n_prime_rel}) and below throughout the text where we sum over momenta, we
set the system volume equal to unity thus omitting a trivial normalization factor.

The genuine chemical potential is then given by
\begin{equation}
\mu = nU - |\tm|\, .
\label{mu_genuine}
\end{equation}
The pressure and entropy density are  obtained as 
\begin{eqnarray}
p\, = \,  \tm^2/2U + 2n'|\tm| + (n')^2 U \nonumber \\
-{\frac 1 2}\sum_{\bf k} \left\{  E(k)-\epsilon(k) -|\tm| - \tm^2 \Pi (k) 
+ 2T\ln \left[1-{\rm e}^{-E(k)/T} \right]
\right\}  , ~\label{pressure}
\end{eqnarray}
and
\begin{eqnarray}
s=\sum_{\bf k}\, \left[{E(k)\over T} N_E -\ln\left(1-{\rm e}^{-E(k)/T}\right)
\right]  , \label{entropy}
\end{eqnarray}
from which the energy density follows from the general relation $\varepsilon=\mu n+Ts-p$.

At zero temperature  the integrals can be evaluated analytically, as discussed in subsection \ref{subsec:practical}.

In continuous space, where $\epsilon(k)=k^2/2m$, the superfluid density is given compactly by Landau's formula
\begin{equation}
n_{\rm s}=n+{1\over d}\sum_{\bf k}\,  \frac{k^2}{m} \frac{\partial N }{ \partial E} \, .
\label{Landau}
\end{equation}
Due to the condition (\ref{dilute}) the answer for a lattice differs  {\it only} by a global factor equal to the ratio between the bare and effective masses.

\subsection{Single-particle density matrix. Quasi-condensate}

In the superfluid region, the single-particle density matrix can be parameterized as
\begin{equation}
\rho({\bf r}) =
\tilde{\rho}({\bf r})\,  {\rm e}^{-\Lambda ({\bf r}) } ,
\label{rho_of_r}
\end{equation}
with
\begin{equation}
\tilde{\rho}({\bf r}) = n-  \sum_{\bf k}\,  \left( 1- {\rm e}^{{\rm i}{\bf k}{\bf r}} \right)  \frac{\epsilon }{E^2}
\left[ {E-\epsilon-|\tm |\over 2}  +(E +|\tm |)N_E \right] ,
\label{tilde-rho}
\end{equation}
\begin{equation}
\Lambda ({\bf r}) =  \frac{|\tm |}{n}\sum_{\bf k}\, \left( 1- {\rm e}^{{\rm i}{\bf k}{\bf r}} \right) 
\frac{E-\epsilon}{2E^2} \left[1+2N_E \right]  \, .
\label{lambda_rel}
\end{equation}
By definition, $\rho(\infty)$ is the condensate density.
The function $\tilde{\rho}(\infty )$ is finite in any dimensions.  In 1D and at finite temperature in 2D, the function $\Lambda({\bf r})$  diverges at $r \to \infty$, consistently with the general
fact of absence of condensate in those systems. Nevertheless, at distances at which the function $\tilde{\rho}({\bf r})$ saturates to its asymptotic value $\tilde{\rho}(\infty)$, the function $\Lambda({\bf r})$
is still much smaller than unity. This characteristic feature of the weakly interacting system allows one to speak of the quasi-condensate with the density given by $\tilde{\rho}(\infty)$. Up to the distances at which $\Lambda$ becomes $\sim 1$,
the correlation properties of condensed and quasi-condensed systems are essentially the same.
Details of the general approach to long-range off-diagonal many-particle correlation functions are described in section \ref{sec:off}.

\subsection{Thermodynamics in the normal region}
The density $n\equiv n(\tm, T)$ in the normal region is given by
\begin{equation}
n\, =\,  \sum_{\bf k} \, \left[{\rm
e}^{\tilde{\epsilon}(k)/T} - 1 \right]^{-1} \, , \label{n_normal_rel}
\end{equation}
with
\begin{equation}
\tilde{\epsilon}(k)\, =\, \epsilon(k) + |\tm|  \; . \label{eff_e_rel}
\end{equation}
The pressure is obtained by
\begin{equation}
p\, = \,n^2 U -T \sum_{\bf k} \, \ln
\left[1-{\rm e}^{-\tilde{\epsilon}(k)/T} \right] \; .
\label{p_normal_rel}
\end{equation}
For the entropy  and energy densities we have the relations
\begin{equation}
s=\sum_{\bf k} \, \left[{\tilde{\epsilon}(k)\over T}N_{\tilde{\epsilon}}
-\ln\left(1-{\rm e}^{-\tilde{\epsilon}(k)/T}\right) \right] \, , \label{s_nrml_rel}
\end{equation}
\begin{equation}
\varepsilon=Un^2+\sum_{\bf k} \, \epsilon (k)N_{\tilde{\epsilon}} \, .
\label{e_nrml_rel}
\end{equation}

\subsection{Accuracy control and fluctuation region}
A necessary condition for the  above-outlined relations  to apply is the smallness
of the parameter
\begin{equation}
 \gamma _0 \; = \;\left\{ {\begin{array}{*{20}c}
  {\sqrt {na^3 } \qquad (d = 3),}  \\
  {mU\qquad \;(d = 2),}  \\
  {\sqrt {mU/n} \qquad (d = 1) .} \\
\end{array}} \right.
\label{gamma_0_summary}
\end{equation}
At $T\lesssim nU$, the parameter $\gamma_0$ directly controls  the systematic error of the theory, and the condition $\gamma_0\ll 1$ is sufficient
for the theory to be accurate. At higher temperatures,  the condition (\ref{gamma_0_summary}) is only  {\it necessary}, since there exists the so-called fluctuation region where 
the fluctuations of the order parameter are essentially non-linear and are not captured by our theory. The closeness to the fluctuation region is described by the dimensionless parameter 
\begin{equation}
x={\mu -\mu_{\rm c}^{(d)}(T) \over (m^d T^2 U^2)^{1\over 4-d} } \, .
\label{param_mu_sum}
\end{equation}
In 2D and 3D systems, $\mu_{\rm c}^{(d)}(T)$ is the critical value of the chemical potential for a given temperature (for the explicit expressions, see section \ref{sec:fluct}), in 1D systems, where there is
no finite-temperature phase transition (see subsection \ref{subsec:specifics_1D} for more detail), $\mu_{\rm c}^{(1)}\equiv 0$.
The theory applies as long as $|x|\gg 1$,   getting progressively less accurate with decreasing  $|x|$. At $|x|\lesssim1$, the theory fails to properly describe condensate and superfluid
densities the values of which being defined by the fluctuating classical-field order parameter. The description of other thermodynamic quantities is better since the fluctuation contributions to
them are of the higher order than the leading (and sometimes even sub-leading) terms. 

The estimates for systematic errors  for major thermodynamic quantities away from and within the fluctuation region are given in sections \ref{sec:Estimates} (in the superfluid region) and \ref{subsec:exp_par_norm} (in the normal region).

In the fluctuation region and its vicinity, the theory can be improved by incorporating accurate description of fluctuation contributions by---universal to all weakly interacting U(1) models
and available in literature---dimensionless scaling functions. The description is outlined in section \ref{sec:fluct}.

\section{Beliaev diagrammatic technique revisited}
\label{sec:Diagrams}

\subsection{The Hamiltonian and approach}

The diagrammatic technique for bosons can be derived in terms of functional integrals over the classical complex-valued fields $\psi({\bf r},\tau )$, propagating in the imaginary time from $\tau =0$ to $\tau = \beta \equiv 1/T$ (we set $\hbar=1$), and subject to the $\beta$-periodicity constraint with respect to the variable $\tau$ \cite{Popov87}.
The classical-field grand-canonical Hamiltonian has a form
\begin{equation}
H\, =\, H_0+H_{\rm int}+H_1\; , \label{H}
\end{equation}
where
\begin{equation}
H_0\, =\, {1\over 2m}\int |\nabla \psi|^2 {\rm d}{\bf r}\;  ,
\label{H_0}
\end{equation}
is the ideal-gas term, with $m$ the particle mass, and
\begin{equation} H_{\rm int}\, =\, {\frac 1 2}\int {\cal U} ({\bf r}_1-{\bf r}_2) |\psi({\bf r}_1 )|^2 |\psi({\bf r}_2 )|^2
{\rm d}{\bf r}_1{\rm d}{\bf r}_2  \; , \label{H_int}
\end{equation}
is the pairwise interaction term. The $H_1$ term contains linear and bilinear terms associated with the chemical potential, $\mu$, external potential, $V({\bf r})$, and the field $\eta({\bf r})$
which explicitly breaks the U(1) symmetry,
\begin{equation}
H_1\, =\, \int [V({\bf r}) -\mu] |\psi|^2 {\rm d}{\bf r} \, - \int
[\eta^*\psi + {\rm c.c.}] \,  {\rm d}{\bf r}  \, . \label{H_1}
\end{equation}
For a homogeneous system $V(r)=0$. Strictly speaking,
at finite $\eta$ one cannot refer to $\mu $ as a
chemical potential because the total number of particles is not
conserved. However, we employ $\eta$-terms solely for explicitly breaking the symmetry
and stabilizing supercurrent states; at the end of the calculation we take the
limit $|\eta|\to 0$. We will therefore continue referring to $\mu$ as the chemical potential.

Standard finite-temperature treatments of the WIBG typically suffer from
infra-red divergencies \cite{Nepom78,Popov79,Nepom83,Griffin98}, especially in lower dimensions.
The ultra-violet divergence is removed by introducing the notion of the pseudopotential
which allows one to express final answers in terms of the scattering length alone.
Our treatment successfully and systematically deals with divergence issues,
and features (at least) three important considerations: \\
- First, the derivation is based on the appropriate definition of the pseudopotential
$U(k) \approx U(0)$ for low momenta [see the discussion of equation (\ref{U_of_k}) below]
in any dimension. \\
- Second, the U(1)-symmetry breaking field $\eta$ introduces a gap in the spectrum of Goldstone modes in a well-controlled way. This suppresses long-range phase fluctuations of the order parameter and removes infra-red divergencies in the behaviour of correlation functions and self-energies without modifying any of the physically important quantities in the relevant order of approximation. In lower dimensions, finite $\eta$ results in a finite
value of the genuine condensate density which dramatically simplifies the description.  In this respect, a small $\eta$ does essentially the same job as Berezinskii's finite-size trick, cf.~\cite{KSS}, or Popov's special (hydrodynamic) treatment of long-wave parts of the fields \cite{Popov87}.
In all final answers one can set $\eta=0$. \\
- Third, the effects of {\it both} interaction and chemical/external potential are fundamentally
non-perturbative separately below the critical temperature. In the absence of interactions a positive chemical potential immediately leads to the instability of the ideal Bose gas, so that there is no way of consistently introducing a positive $\mu$---crucial for describing the system below the critical point---before switching on the interaction. This observation explains the idea behind omitting
all terms in (\ref{H_1}) from the non-interacting Hamiltonian, $H_0$. At the end of the day,
we will use $\mu - 2nU(0)$ and $T$ as two variables describing the thermodynamic ensemble.

\subsection{Diagrammatic expansion}

\begin{figure}
\includegraphics[angle=-90, scale=0.33, bb=180 -250 380 500]{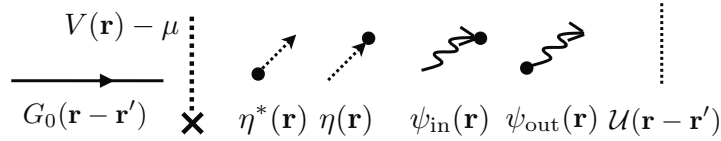}
\caption{Graphical objects representing the single particle
propagator $G^{(0)}({\bf r}-{\bf r}')$, the external field $V({\bf r})-\mu$ , the symmetry breaking fields $\eta^*(\bf{r})$ and $\eta(\bf{r})$, the condensate lines $\psi_{\rm in}(\bf{r})$ and $\psi_{\rm out}(\bf{r})$, and the  interaction ${\mathcal{U}}({\bf r} -  {\bf r}')$ from left to right. Here, and in all other figures in the paper, we will only write down the coordinates or the momenta of the diagrammatic elements, omitting the frequency or imaginary time dependence.}
\label{fig:Hamiltonian}
\end{figure}

Since the non-interacting Hamiltonian corresponds to the ideal gas at
chemical potential approaching zero from below, i.e. with no Bose-Einstein condensate,
the corresponding (bare) Matsubara Green's function reads
\begin{eqnarray}
G^{(0)}(\tau_1, \tau_2,{\bf r}_1, {\bf r}_2 ) =
\sum_{\xi, {\bf k}}  G^{(0)}(\xi,{\bf k})\,  {\rm e}^{{\rm i}{\bf k}\cdot ({\bf r}_1- {\bf r}_2 )-{\rm i}\xi (\tau_1-\tau_2)} \,  , \label{G_0_r}
\end{eqnarray}
\begin{equation}
G^{(0)}(\xi,{\bf k} )\, =\, \left[ {\rm i} \xi - \epsilon(k) \right]^{-1}\, ,
\label{G_0_k}
\end{equation}
where $\epsilon(k)$ is the single particle dispersion relation. We typically assume
a parabolic dispersion relation $\epsilon(k)=k^2/2m$, but our final answers
(excluding formulae for superfluid properties which explicitly invoke Galilean invariance)
are valid for
arbitrary $\epsilon(k)$ with a parabolic dependence on momentum in the long-wave limit,
e.g.\ for the tight-binding spectrum.
We use the Matsubara imaginary-frequency representation:
$\xi \equiv \xi_s = 2s\pi T$ ($s=0,\pm 1, \pm 2 , \ldots$) with $T$ the temperature.
For graphical representation of diagrammatic expansions and relations we
introduce a set of objects in figure~\ref{fig:Hamiltonian} depicting the single-particle
propagator, the condensate, and various terms in the Hamiltonian.

\begin{figure}
\includegraphics[angle=-90, scale=0.33, bb=100 -220 420 500]{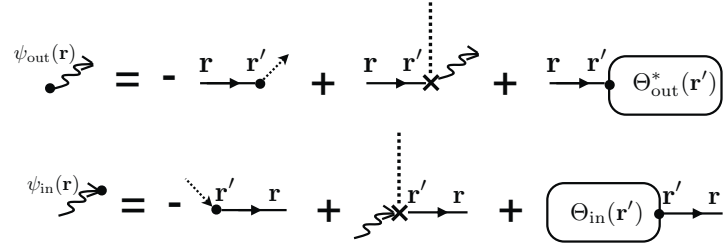}
\caption{Diagrammatic expansion for (\ref{psi_in}) and (\ref{psi_out}). }
\label{fig:Dyson_psi}
\end{figure}

The non-perturbative response to $H_{\rm int}$ and $H_1$ is
accounted for by Dyson summation. First, we consider diagrams
that feature only one incoming or outgoing line---we call them tails.
Given our starting point with no condensate in
the non-perturbed system such particle-number-changing diagrams exist only due to the symmetry
breaking field $\eta$. The frequency of the line connecting $\eta$ to the rest of the
diagram is zero, by frequency conservation.
The Dyson summation of all tails attached to a given point replaces them with a single line which
we denote as $\psi_{\rm in}({\bf r})$ and $\psi_{\rm out}({\bf
r})$, if the tails are incoming and outgoing, respectively, see figure~\ref{fig:Dyson_psi}. Below we write expressions in the frequency representation; if the frequency 
argument is not mentioned it is implied that
its value is zero. We also adopt a convention of integration over repeated
coordinate/momentum/frequency arguments. The
Dyson equation for $\psi_{\rm in}({\bf r})$ then reads:
\begin{eqnarray}
\psi_{\rm in}({\bf r})\, =\, -G^{(0)}({\bf r}- {\bf r}' )\eta({\bf
r}') + 
\nonumber \\
G^{(0)}({\bf r}- {\bf r}' )[V({\bf r}') -\mu]\psi_{\rm
in}({\bf r}')+G^{(0)}({\bf r} - {\bf r}' )\Theta_{\rm in}({\bf r}')\, . \label{psi_in}
\end{eqnarray}
Here $\Theta_{\rm in}$ is the sum of all other diagrammatic elements
attached to the first line which are not accounted for by the first two terms, i.e. excluding
diagrams with the field $\eta$ and diagrams connected to the first solid line by the
$[V({\bf r}) -\mu]$ vertex. The subscript `in' reminds that $\Theta_{\rm in}$ has an extra
incoming particle line.
Similarly,
\begin{eqnarray}
\psi_{\rm out}({\bf r})\, =\, -\eta^*({\bf r}')G^{(0)}({\bf r}'-
{\bf r} )   +  \nonumber \\
\psi_{\rm out}({\bf r}')[V({\bf r}') -\mu]G^{(0)}({\bf
r}'- {\bf r} ) +   \Theta_{\rm out}({\bf r}')G^{(0)}({\bf r}'- {\bf r} ) \, . 
\label{psi_out}
\end{eqnarray}
The fact that $G^{(0)}({\bf r} )$ is a real even function of its argument implies that
\begin{equation}
\psi_{\rm out}\, =\, \psi_{\rm in}^*\, , \qquad \Theta_{\rm out}\,
=\, \Theta_{\rm in}^*  \; . \label{conj_1}
\end{equation}

The diagrammatic expansion for the tail is identical to that for the condensate wave function
defined as an anomalous average
\begin{equation}
\psi_0({\bf r})\ =\ \langle \psi ({\bf r}) \rangle \ \equiv \  \psi_{\rm in}({\bf r}) \; .
\label{cond_w_f}
\end{equation}
Correspondingly, the condensate density is defined as $n_0 ({\bf r}) = \vert \psi_0({\bf r}) \vert^2$.
From now on we will write $\psi_0$ for $\psi_{\rm in}$ and $\psi_0^*$ for $\psi_{\rm out}$, and use the notions of condensate lines and tails on equal footing. Of course, in the limit of $\eta \to 0$
speaking of the condensate wave-function is meaningful only when the long-range fluctuations
of the phase are small.

\subsection{Normal and anomalous propagators}

\begin{figure}
\includegraphics[angle=-90, scale=0.2, bb=150 -700  400 -100]{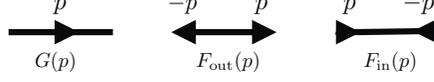}
\caption{Symbols used for normal, $G$, and anomalous, $F$,  propagators in a homogeneous system.}
\label{fig:propagators}
\end{figure}

\begin{figure}
\includegraphics[angle=-90, scale=0.33, bb=-50 -250 570 200]{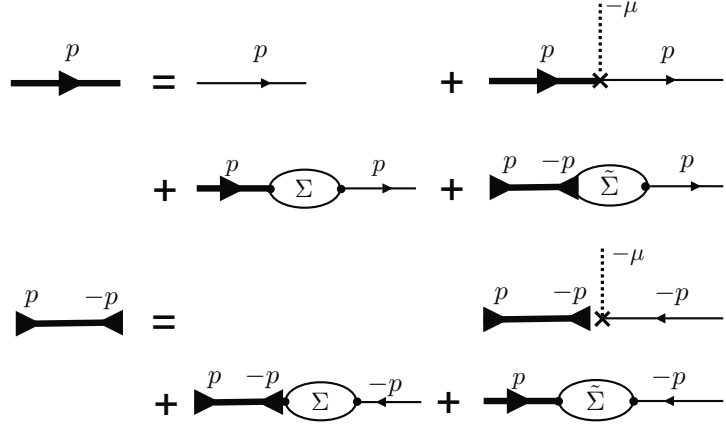}
\caption{Beliaev-Dyson equations for a homogeneous system.}
\label{fig:Beliaev-Dyson}
\end{figure}

The next step is to introduce exact, or  `bold', particle propagators and work with skeleton diagrams.
There are a couple of differences between our approach and the standard Beliaev technique. Our
bare (thin-line) propagators have zero chemical potential and this, according to (\ref{psi_in}), immediately results in a constraint relating the condensate density to the chemical potential and
$\Theta_{\rm in}$ [see (\ref{chem_pot_rel}) below]. Also, the chemical/external potential have to be explicitly introduced into the standard Beliaev-Dyson equations for the normal and anomalous
Green's functions, see e.g.,~\cite{AGD, FetterWalecka, Griffin98}. These equations
can be defined purely diagrammatically. To this end---proceeding in the frequency representation for the sake of definiteness---we introduce the normal Green's function, $G(\xi,
{\bf r}_1, {\bf r}_2)$, defined as the sum---with a global minus sign,
which is a mere convention---of all diagrams which have an
incoming $G^{(0)}$-line with frequency $\xi$ to point ${\bf
r}_1$  and an outgoing $G^{(0)}$-line (with the same frequency)
from point ${\bf r}_2$. The anomalous Green's
function, $F_{\rm in}(\xi, {\bf r}_1, {\bf r}_2)$, by definition,
has an incoming $G^{(0)}$-line with frequency $\xi$ to point
${\bf r}_1$ and another incoming $G^{(0)}$-line with
frequency $-\xi$, by conservation of frequency, to point ${\bf
r}_2$. The anomalous Green's function $F_{\rm out}(\xi, {\bf r}_1,
{\bf r}_2)$ is a counterpart of the function $F_{\rm in}(\xi, {\bf
r}_1, {\bf r}_2)$: instead of two incoming it has two outgoing
$G^{(0)}$-lines; one with frequency $\xi$ from point
${\bf r}_1$ and another with
frequency $-\xi$ from point ${\bf r}_2$.
The symmetry under exchanging the end points of the anomalous Green's
functions immediately implies the following relations
\begin{equation}
F_{\rm in}(\xi, {\bf r}_1, {\bf r}_2)\, =\, F_{\rm in}(-\xi, {\bf
r}_2, {\bf r}_1)  \; , \label{sym1}
\end{equation}
\begin{equation}
F_{\rm out}(\xi, {\bf r}_1, {\bf r}_2) = F_{\rm out}(-\xi,
{\bf r}_2, {\bf r}_1)  \; . \label{sym2}
\end{equation}
Since complex conjugation is equivalent to changing the sign of the Matsubara frequency
and direction of propagation we also have
\begin{equation}
[G(\xi, {\bf r}_1, {\bf r}_2)]^*\, =\, G(-\xi, {\bf r}_2, {\bf
r}_1) \;,  \label{sym3}
\end{equation}
\begin{equation}
F_{\rm in}^*(\xi, {\bf r}_1, {\bf r}_2) = F_{\rm out}(-\xi,
{\bf r}_2, {\bf r}_1) =F_{\rm out}(\xi, {\bf r}_1, {\bf r}_2)
\;. \label{sym4}
\end{equation}

The physical meaning of $G$, $F_{\rm in}$, and $F_{\rm out}$ follows from the structure
of the two-point correlation functions in the imaginary-time--coordinate representation:
\begin{eqnarray}
\langle  \psi(\tau_1, {\bf r}_1)  \psi^* (\tau_2, {\bf r}_2) \rangle
=  - G(\tau_1-\tau_2, \,  {\bf r}_1, {\bf r}_2) + \psi_0 ({\bf r}_1)\psi_0 ^*({\bf r}_2) ,~~~
\label{G_phys}
\end{eqnarray}
\begin{eqnarray}
\langle  \psi(\tau_1, {\bf r}_1)  \psi (\tau_2, {\bf r}_2) \rangle
=  F_{\rm in} (\tau_1-\tau_2, \,  {\bf r}_1, {\bf r}_2) + \psi_0 ({\bf r}_1)\psi_0 ({\bf r}_2) .~~~
\label{F_in_phys}
\end{eqnarray}
These relations can be readily checked by expanding the averages into diagrammatic
series. The special case of (\ref{G_phys}), corresponding to $\tau_1 \to \tau_2 -0$ and ${\bf r}_1={\bf r}_2$, relates the local density to 
the normal Green's function and the condensate density:
\begin{equation}
n({\bf r})
=  - G(\tau=-0,  {\bf r}, {\bf r}) +n_0({\bf r}) \; .
\label{G_phys_special}
\end{equation}

The Beliaev-Dyson equations then read
\begin{eqnarray}
G(\xi,{\bf r}_1,{\bf r}_2)\, =\, G^{(0)}(\xi,{\bf r}_1,{\bf r}_2)
+ \, G(\xi,{\bf r}_1,{\bf r}')[V({\bf r}')-\mu]G^{(0)}(\xi,{\bf
r}',{\bf r}_2) \nonumber \\ 
+G(\xi,{\bf r}_1,{\bf r}')\, \Sigma_{11}(\xi,{\bf r}',{\bf r}'')\, G^{(0)}(\xi,{\bf r}'',{\bf
r}_2) \nonumber \\ 
+\, F_{\rm in}(\xi,{\bf r}_1,{\bf r}')\, \Sigma_{20}(\xi,{\bf r}',{\bf r}'')\, G^{(0)}(\xi,{\bf r}'',{\bf r}_2)
\, , \label{DB1}
\end{eqnarray}
\begin{eqnarray}
F_{\rm in}(\xi,{\bf r}_1,{\bf r}_2)\, =\, F_{\rm in}(\xi,{\bf
r}_1,{\bf r}')[V({\bf r}')-\mu]G^{(0)}(-\xi,{\bf r}_2,{\bf r}')
\nonumber \\
+ \, F_{\rm in}(\xi,{\bf r}_1,{\bf r}')\, \Sigma_{11}(\xi,{\bf r}',{\bf r}'')\, G^{(0)}(-\xi,{\bf r}_2,{\bf r}'') \nonumber \\ 
+\, G(\xi,{\bf r}_1,{\bf r}')\, \Sigma_{02}(\xi,{\bf r}',{\bf r}'')\, G^{(0)}(-\xi,{\bf r}_2,{\bf r}'') \, , \label{DB2}
\end{eqnarray}
with the standard definition of self-energies $\Sigma$'s as sums of diagrams
which can not be cut through a single $G$ or $F$ line. Equations~(\ref{DB1}) and (\ref{DB2})
are shown graphically in figure~\ref{fig:Beliaev-Dyson} for a homogeneous system
in momentum representation. The complexity of the theoretical solution is in
the evaluation of the $\Theta$ and $\Sigma$ functions.

\subsection{The chemical potential and the Hugenholtz-Pines relation}

Since the bare Green's function at zero frequency is identical to the inverse Laplacian
operator one can cast (\ref{psi_in}) in the differential form
\begin{equation}
-{\Delta \over 2m}\, \psi_0({\bf r}) +[V({\bf r}) -\mu]\,
\psi_0({\bf r})+ \Theta_{\rm in}({\bf r}) \, =\, \eta({\bf r})\; .
\label{psi_in_dif}
\end{equation}
This equation reduces to the Gross-Pitaevskii equation at low enough temperature
when the leading term in $\Theta_{\rm in}$ is $\propto \vert \psi_0({\bf r}) \vert^2 \psi_0({\bf r})$.
In the homogeneous case at $\eta\to 0$, when $\psi_0$ and $\Theta_{\rm
in}$ are coordinate-independent, equation (\ref{psi_in_dif}) simplifies to
\begin{equation}
\mu\,  =\,  \Theta_{\rm in}/\psi_0  \equiv \Theta_{\rm  out} / \psi_0^* , \label{chem_pot_rel}
\end{equation}
generalizing relations (21.3)-(21.5), and $\mu = \langle 0 \vert \partial (H_{\rm int}/V) /\partial n_0 \vert 0 \rangle$,  in \cite{FetterWalecka} to finite temperatures.

There is also an exact relation between $\Sigma_{11}$, $\Sigma_{02}$,
$\Theta_{\rm in}$, and $\psi_0$ (see also \cite{Pines}).
Here (and only here!) we assume that all diagrams for $\Sigma$'s are in terms
of $G^{(0)}$ and condensate lines. Let $D^{(l)}_{\rm in}$ be the sum of diagrams
contributing to $\Theta_{\rm in}$ with $l$ incoming and $l-1$ outgoing condensate lines.
Then (for $\xi=0$)
\begin{equation}
\Sigma_{11}({\bf r}, {\bf r}')\, \psi_0 ({\bf r}')\, =\,
\sum_{l=1}^{\infty}\, l D^{(l)}_{\rm in}
\; , \label{rel_sigma_11}
\end{equation}
because each diagram, with $l-1$ incoming condensate lines, contributing to $\Sigma_{11}({\bf r}, {\bf r}')$ produces---upon integration over ${\bf r}'$ with the weight
$\psi_0 ({\bf r}')$---a diagram contributing to $D^{(l)}_{\rm
in}$, and there are $l$ such diagrams contributing to $\Sigma_{11}$.
An identical argument leads to
\begin{equation}
\Sigma_{02}({\bf r}, {\bf r}')\, \psi_0^* ({\bf r}')\, =\,
\sum_{l=2}^{\infty}\, (l-1) D^{(l)}_{\rm in}
\; . \label{rel_sigma_02}
\end{equation}
By subtracting (\ref{rel_sigma_02}) from (\ref{rel_sigma_11})
we obtain
\begin{equation}
\Sigma_{11}( {\bf r}, {\bf r}')\, \psi_0 ({\bf
r}')-\Sigma_{02}( {\bf r}, {\bf r}')\, \psi_0^* ({\bf r}')\, =\,
\Theta_{\rm in}({\bf r})  \; . \label{combin}
\end{equation}

In the homogeneous case $\psi_0 = \psi_0^*$. Then,
in the $\eta \to 0$ limit, equations(\ref{combin}) and (\ref{chem_pot_rel})
can be combined to yield the Hugenholtz-Pines relation~\cite{Pines} (and its
finite-temperature version~\cite{Hohenberg65})
\begin{equation}
\mu\, =\, \Sigma_{11}(0,0)-\Sigma_{02}(0,0)
\; . \label{HP}
\end{equation}

\subsection{Beliaev-Dyson equations in the presence of homogeneous superflow}

In order to discuss the superfluid properties of the homogeneous system, we add a phase  factor
to the symmetry breaking field
\begin{equation}
\eta({\bf r})=\eta_0\, {\rm e}^{{\rm i}{\bf k}_0\cdot {\bf r}} \; .
\label{eta_flow}
\end{equation}
which readily translates into phase of the condensate wave function
\begin{equation}
\psi_0({\bf r})=\sqrt{n_0}\, {\rm e}^{{\rm i}{\bf k}_0\cdot {\bf r}} \;,
\label{flow}
\end{equation}
The only difference with the previous discussion is that now
we have to associate a finite momentum $\pm {\bf k}_0$ carried by the condensate lines
and modify the momentum conservation laws accordingly. Then
(\ref{chem_pot_rel}) and (\ref{HP}) become
\begin{equation}
\mu\, =\, \Theta/\sqrt{n_0}+k_0^2/2m-\eta_0/\sqrt{n_0}
\; , \label{1}
\end{equation}
\begin{equation}
\Sigma_{11}({\bf k}_0) - \Sigma_{02}(0)  = \frac{\Theta}{\sqrt{n_0}} = \mu-\frac{k_0^2}{2m} +
\frac{\eta_0}{\sqrt{n_0}} \; ,
\label{2}
\end{equation}
where $\Theta_{\rm in} = \Theta_{\rm out} \equiv \Theta$.

It is convenient (for transparency of expressions which follow) to combine
frequency and momentum into a single ``4-momentum'' variable $P=(\xi, {\bf k})$ and to introduce an
auxiliary momentum $P'=(-\xi,2{\bf k}_0-{\bf k})$.
The symmetry between the two ends of the anomalous Green's functions
and, equivalently, between the two ends of the anomalous self-energies
is then expressed by (accounting for the momentum carried by the condensate lines)
\begin{equation}
F_{\rm in/out}(P)\, =\, F_{\rm in/out}(P')
\; , \label{F_sym_mom}
\end{equation}
\begin{equation}
\Sigma_{20/02}(P)\, =\, \Sigma_{20/02}(P')
\; . \label{Sigma_sym_mom}
\end{equation}
[In a more comprehensive notation scheme one has to mention momenta
of both incoming lines in $F_{\rm in/out}(P,P')$.]

Complex conjugation of propagators and condensate lines changes the signs of their
4-momenta. This property can be used to prove the symmetry relation
for the Green's function
\begin{equation}
G^*( P )\vert_{ {\bf k}_0 }  = G(-P )\vert_{ -{\bf k}_0} \; .
\label{G_P_sym}
\end{equation}
Similar symmetry relations take place for the anomalous Green's
functions and all three self-energies.

In the momentum representation, inverting the direction of ${\bf k}_0$
does not change the analytical expression for propagators. Hence, by
inverting the direction of all the lines, including the condensate
ones, it follows that
\begin{equation}
F_{\rm in}(P) \, =\, F_{\rm out}(P) \, \equiv \, F(P) \; ,
\label{F_P_sym}
\end{equation}
\begin{equation}
\Sigma_{20}(P) \, =\, \Sigma_{02}(P) \, \equiv \,
\tilde{\Sigma}(P)\; . \label{Sigma_20_P_sym}
\end{equation}
(For the normal Green's function, $G(P)$, and the self-energy, $\Sigma_{11}(P)$,
inverting momentum directions results in the same series, and in this sense is trivial.)

We are ready to formulate the pair of Beliaev-Dyson equations
in the momentum representation. Using symmetry properties
into consideration and the shorthand notation $\Sigma\equiv \Sigma_{11}$
we obtain
\begin{eqnarray}
G(P) =
G^{(0)}(P)+G(P)[\Sigma(P)-\mu]G^{(0)}(P) \nonumber \\
+F(P)\tilde{\Sigma}(P)G^{(0)}(P)\, ,   \\ \label{DB1_P}
F(P) = F(P)[\Sigma(P')\! -\! \mu]G^{(0)}(P') 
 + G(P)\tilde{\Sigma}(P)G^{(0)}(P')\, . 
\label{DB2_P}
\end{eqnarray}
The solution in terms of self-energies reads
\begin{equation}
G(P)\, =\, {{\rm i}\xi+\epsilon(|2{\bf k}_0 - {\bf k}|) + \Sigma(P')-\mu
\over D(P)}\; , \label{G_expr}
\end{equation}
\begin{equation}
F(P)\, =\, -{\tilde{\Sigma}(P) \over D(P)}\; , \label{F_expr}
\end{equation}
where
\begin{equation}
D = \tilde{\Sigma}^2(P) - [\epsilon(|2{\bf k}_0 - {\bf k}|) +
\Sigma(P')-\mu +{\rm i}\xi][\epsilon(k) + \Sigma(P)-\mu -{\rm i}\xi]  .~
\label{denom}
\end{equation}
With these relations at hand, one can calculate the current density induced by
the phase gradient in the condensate wave-function, see subsection \ref{subs:SD}.

\subsection{Low-density limit in 3D and 2D: Pseudo-potentials and scattering lengths }

\begin{figure}
\includegraphics[angle=-90, scale=0.25, bb=100 -400  450 200]{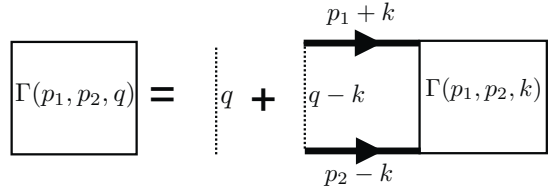}
\caption{
Ladder diagrams leading to  equation (\ref{eq:Gamma}). }
\label{fig:ladder}
\end{figure}

\begin{figure}
\includegraphics[angle=-90, scale=0.2, bb=100 -600  450 0]{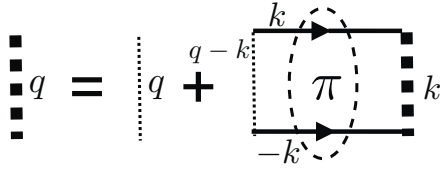}
\caption{The diagrammatic expression for the pseudo-potential (left hand side) involves the bare potential 
(first term on the right hand side) and a modified two-particle propagator $\Pi(\bf {k})$. }
\label{fig:pseudo_potential}
\end{figure}

In two and three dimensions, the expansion in terms of the bare interaction potential can be (and in most realistic cases, is) non-perturbative.
The system is regarded as weakly-interacting only because of the low density of particles.
In one dimension, the physics is perturbative instead in the high-density limit for a given potential. Theoretically, dealing with the strong bare potential implies summation of an infinite
sequence of ladder diagrams and produces an effective interaction in the form of the four-point vertex~\cite{Beliaev}, $\Gamma$, see figure~\ref{fig:ladder}.  The analytical relation behind figure~\ref{fig:ladder} reads
\begin{eqnarray}
\Gamma (P_1,P_2,Q)\, =\,\nonumber \\
 {\cal U}({\bf q})+\sum_{K}\, {\cal U}({\bf q}-{\bf k})\, G(P_1\! +\! K)\, G(P_2\! -\! K)\, \Gamma (P_1,P_2,K) \label{eq:Gamma}  \\
\equiv  \, {\cal U}({\bf q})+\sum_{K}\, \Gamma(P_1\! +\! K,P_2\! -\! K,Q-K) \, G(P_1\! +\! K)\, G(P_2\! -\! K)\,  {\cal U}({\bf k}) \nonumber \,  .
\end{eqnarray}
When the bare-interaction lines are replaced with $\Gamma$'s, the rest of the
series becomes perturbative (excluding the critical region).
On the technical side, working with $\Gamma$'s is convenient
as long as one does not intend to systematically take into account higher-order corrections,
utilizing thus only the (simple and transparent) leading-order expression for $\Gamma (0,0,0)$.
For the higher-order corrections, equation (\ref{eq:Gamma}) involves three different length scales:
the size of the potential, $R_0$, the healing length of the Bose condensed system, and the de Broglie wavelength, while only the non-perturbative physics at the scale $R_0$ requires summation of the ladder diagrams. Hence, it makes sense to construct an object with a much simpler structure than $\Gamma$ that captures the non-perturbative physics (and thus coincides with the leading-order expression for
$\Gamma (0,0,0)$), and then systematically investigate the difference between this object and $\Gamma$.
To achieve this goal let us define the {\it pseudo-pontential} $U({\bf q})$ by the equation  (see figure~\ref{fig:pseudo_potential})
\begin{eqnarray}
U({\bf q})\, =\, {\cal U}({\bf q})+\sum_{\bf k }\, {\cal U}({\bf q}-{\bf k})\,  \Pi ({\bf k}) \, U({\bf k}) \label{eq:def_of_U} \\
\equiv \, {\cal U}({\bf q})+\sum_{\bf k }\, U({\bf q}-{\bf k})\,  \Pi ({\bf k}) \, {\cal U}({\bf k}) \nonumber \; ,
\end{eqnarray}
where $\Pi ({\bf k})$ is such that when $k$ is much larger than the inverse healing length or thermal momentum the value of  $\Pi ({\bf k})$ approaches that of the product $G(P_1\! +\! K)G(P_2\! -\! K)$ summed over the frequency of the 4-vector $K$. That is,
\begin{equation}
\Pi ({\bf k})  \, \to \, -{1\over 2\epsilon(k)} \quad {\rm at}\quad k\, \to\, \infty \, .
\label{Pi}
\end{equation}
In this case, $\Gamma (P_1,P_2,Q)\, \approx \, U({\bf q})$, and one can expand the difference in a perturbative series.
As a result, we arrive at the diagrammatic technique where instead of thin dashed lines standing for the bare interaction potential we have bold dashed lines representing
the pseudo-potential $U$, with an additional requirement that whenever two (normal) Green's function lines are sandwiched between two pseudo-potential lines, the former
are  supposed to be summed over the frequency difference (which is always possible in view of the  frequency-independence of the pseudo-potential), and then $\Pi$ has to be subtracted
from the result of summation. Specifically, if $P_1$ and $P_2$ are the two external 4-momenta of the
above mentioned diagrammatic element, then the following replacement is supposed to take place
for internal propagator lines
\begin{eqnarray}
\sum_{\xi^{(K)}} G(P_1\! +\! K)G(P_2\! -\! K)\, \to\, 
\sum_{\xi^{(K)}} G(P_1\! +\! K)G(P_2\! -\! K) - \Pi({\bf k}) \,  ,
\label{corrected}
\end{eqnarray}
where $\xi^{(K)}$ is the frequency of the 4-momentum $K$.


As long as we are not interested in the non-universal ultraviolet corrections,
we can replace  $U({\bf q})$ with $U(0)$, the systematic error introduced by the replacement being
controlled by the following dimensional estimate
\begin{equation}
U({\bf q})\, =\, U(0)\left[ 1\, +\, {\cal O}(q^2R_0^2) \right]\, .
\label{U_of_k}
\end{equation}
One may wonder how this estimate is reconciled with the momentum
dependence of $\Gamma$, which is of the order $qR_0$ in 3D, and of the order
$1/\ln (1/ qR_0)$ in 2D. The solution is provided by (\ref{corrected})
explaining that this dependence, which is both universal and perturbative, is
taken into account by the second-order ladder-type diagram in $U$.

In 3D, a natural choice for $\Pi({\bf k})$ is
\begin{equation}
\Pi ({\bf k})  \, = \, -{1\over 2 \, \epsilon(k)} \qquad (d=3) \,  ,
\label{Pi_3D}
\end{equation}
in which case we have the usual~\cite{AGD, FetterWalecka} (see also below)
\begin{equation}
U(0)  \, = \, {4\pi a\over m} \qquad (d=3) \,  ,
\label{U_3D}
\end{equation}
with $a$ the $s$-wave scattering length.
In 2D one cannot use (\ref{Pi_3D}) because of the infra-red divergence of the integral in (\ref{eq:def_of_U}). A reasonable choice here is
\begin{equation}
\Pi ({\bf k})  \, = \, -{1\over 2} \, {1\over \epsilon(k) + \epsilon_0}\qquad (d=2) \,  ,
\label{Pi_2D}
\end{equation}
where $\epsilon_0$ is an arbitrary low-energy cutoff. The particular value of $\epsilon_0$ is
not important since final answers are not sensitive to it. Within the systematic-error bars of the pseudo-Hamiltonian description discussed below, any choice of $\epsilon_0$ such that
$nU \lesssim \epsilon_0 \lesssim n/m$ is equally reasonable in terms of accuracy,
provided the temperature is not much larger than $n/m$; moreover, replacing $\epsilon_0$  with
$ck$ is also acceptable. (An optimal choice of $\epsilon_0$   in the regime $T\gg n/m$ will be discussed in subsection \ref{subsec:2D_large_T}.) There is also no need to introduce $\Pi$ in $d=1$.
We will assume that formally $\Pi ({\bf k}) \equiv 0$ in $d=1$ in which case our final answers
can be used as written in all spatial dimensions.

Given that (\ref{Pi_3D}) and (\ref{Pi_2D}) are not unique [there is a free parameter in (\ref{Pi_2D})], it is instructive to explicitly relate two pseudo-potentials, $U_1$ and $U_2$---corresponding to $\Pi_1$ and $\Pi_2$, respectively---to {\it each other}.  We notice that
(\ref{eq:def_of_U}) implies
\begin{eqnarray}
U_2({\bf q}) =U_1({\bf q}) \!+\!
\sum_{\bf k} U_1({\bf q}\! -\! {\bf k})[ \Pi_2 ({\bf k})\! -\! \Pi_1 ({\bf k})] U_2({\bf k}) .~
\label{rel_U1_U2}
\end{eqnarray}
In view of (\ref{U_of_k}) this simplifies [up to  ${\cal O}(q^2R_0^2)$ terms that we neglect in what follows] to
\begin{eqnarray}
U_2=U_1 + U_1C_{12}U_2  \, ,
\label{U1_U2}
\end{eqnarray}
\begin{eqnarray}
C_{12}=\sum_{\bf k} \, [ \Pi_2 ({\bf k})-\Pi_1 ({\bf k})] \,  .
\label{C_12}
\end{eqnarray}
If $\Pi_1$  and $\Pi_2$ are defined by (\ref{Pi_2D}) with different cut-offs $\epsilon_{0}^{(1)}$ and $\epsilon_{0}^{(2)}$,
we have
\begin{equation}
C_{12}  \, =\,  {m\over 4\pi}\, \ln {\epsilon_{0}^{(2)}\over \epsilon_{0}^{(1)}}\qquad (d=2) \,  .
\label{C_12_2D}
\end{equation}
The right choice of the functions $\Pi_1$ and  $\Pi_2$ implies that $U_1$ and $U_2$ differ only by  sub-leading terms. Thus, we can expand the r.h.s. of (\ref{U1_U2}) in powers of $|U_1C_{12}|\ll 1$:
\begin{eqnarray}
U_2=U_1 + U_1^2C_{12} + \ldots \,  .
\label{U1_U2_expd}
\end{eqnarray}

Formally, the expansion in terms of the pseudo-potential is perturbative only as long as we
exclude the high-momentum contributions to the $(M>3)$-body diagrams generating, upon complete summation, $M$-body scattering amplitudes. In a dilute gas, the corresponding diagrams are small
(contain extra powers of the gas parameter $na^3$) and are neglected in this manuscript.

The expression (\ref{U_3D}) has the meaning of mapping a dilute three-dimensional system with an arbitrary short-range interaction potential onto a system with the interaction potential (\ref{hard_pot}), so that the pseudo-potentials of the two systems coincide. The same approach is possible (and popular) in 2D,
the parameter $a$  being called the two-dimensional scattering length, since the mapping
applies to scattering properties as well. For a given potential ${\cal U}$, the value of $a$
can be obtained either from the asymptotic behaviour of the pseudo-potential $U$ at appropriately small $\epsilon_0$, or from the asymptotic behaviour of the scattering amplitude $f({\bf k}, {\bf k}')$ at $k,k'\to 0$ ~\cite{FetterWalecka}. Both asymptotic behaviours are closely related to each other since the large-$k$ limit of the kernel $\Pi ({\bf k})$ in (\ref{eq:def_of_U}) coincides with the large-$k$ limit of the scattering kernel
\begin{equation}
\Pi_{\rm sc} (k,k')  \, ={m\over k'^2-k^2+{\rm i}0}  \,
\label{scatt_krnl}
\end{equation}
in the integral equation for $f({\bf k}, {\bf k}')$,
\begin{eqnarray}
f({\bf k}, {\bf k}') = {\cal U}({\bf k}\! - \! {\bf k}')+
\sum_{{\bf q}} \, {\cal U}({\bf k}\! - \! {\bf q})  \Pi_{\rm sc} (q ,k')\, f({\bf q}, {\bf k}') ~~\label{eq:def_of_t} \\
\equiv  {\cal U}({\bf k}\! - \! {\bf k}')+  \sum_{{\bf q}} \,  f({\bf k}, {\bf q})  \Pi_{\rm sc} (q ,k')\,  {\cal U}({\bf q}\! - \! {\bf k}')\,  . \nonumber
\end{eqnarray}
Moreover, a direct relationship between $U$ and $f({\bf k}, {\bf k}')$ is readily obtained by noticing that (\ref{eq:def_of_t}) and (\ref{eq:def_of_U})  imply [cf. (\ref{rel_U1_U2})]
\begin{eqnarray}
f({\bf k}, {\bf k}')  =  U({\bf k}\! -\!  {\bf k}')+
\sum_{{\bf q}} U({\bf k}\! - \! {\bf q})  [ \Pi_{\rm sc} (q ,k')\! -\! \Pi (q)] f({\bf q} , {\bf k}')\, , 
\label{rel_U_t}
\end{eqnarray}
which dramatically simplifies upon replacing $U({\bf q})$ with $U(0)$, in accordance with (\ref{U_of_k}). In this case, $f({\bf k}, {\bf k}')$ is ${\bf k}$-independent, and for $f\equiv f(k')$
we get
\begin{eqnarray}
f(k') = U+U f(k') \sum_{\bf q} \,  [ \Pi_{\rm sc} (q,k')-\Pi (\bf{q})]  \, .
\label{U_f_smpl}
\end{eqnarray}
Substituting (\ref{Pi_2D}) and (\ref{scatt_krnl}) into (\ref{U_f_smpl}) and performing the integral, we find
\begin{eqnarray}
f(k') = {2\pi/m \over \ln (\sqrt{2m\epsilon_0} / k') + 2\pi /mU+{\rm i}\pi/2} \,  .
\label{eq:tmatrix}
\label{U_f_fin}
\end{eqnarray}
Comparing this to the known hard-disk result (see, e.g., \cite{Mora})
\begin{eqnarray}
f(k') = {2\pi/m \over \ln (2 / ak') -\gamma+{\rm i}\pi/2} \,  ,
\label{t_h_disk}
\end{eqnarray}
where $\gamma=0.5772\dots$ is Euler's constant, we conclude that
\begin{eqnarray}
U = {4\pi/m \over \ln (2/a^2m\epsilon_0) -2\gamma} \,  .
\label{U_a}
\end{eqnarray}
In 3D, equation (\ref{U_f_smpl})---with  $\Pi({\bf{k}})$ given by (\ref{Pi_3D})---yields
\begin{equation}
f(k') = \frac{U}{1- {\rm i} m k' U/4 \pi},
\end{equation}
thus leading to (\ref{U_3D}) by comparison with known expression for the hard-sphere 
scattering amplitude.

\section{Thermodynamic functions in the low-temperature region}
\label{sec:Quasicondensate}

\subsection{Basic relations and notions}
\label{subsec:basic}

\begin{figure}
\includegraphics[angle=-90, scale=0.275, bb=30 -350  510 250]{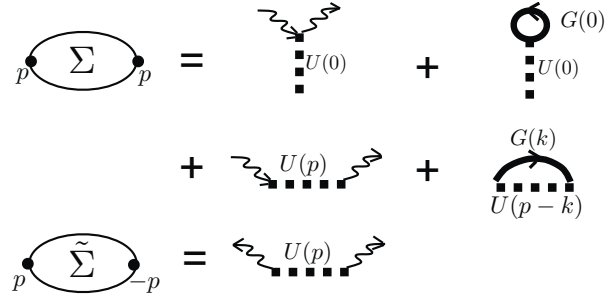}
\caption{The lowest order diagrams for the self-energy $\Sigma$ in (\ref{Sigma_dgr}) and the anomolous seflenergy $\tilde{\Sigma}$ in (\ref{Sigma_anom__dgr}). }
\label{Fig:Sigma11}
\end{figure}

Explicitly calculating the lowest-order diagrams shown in figure~\ref{Fig:Sigma11} and utilizing (\ref{G_phys_special})  one finds
\begin{equation}
\Sigma(P)\, =\, -2G(r=0,\tau=-0)U+2n_0U\, =\, 2nU \; ,
\label{Sigma_dgr}
\end{equation}
\begin{equation}
\tilde{\Sigma}(P)\, =\, n_0U\; . \label{Sigma_anom__dgr}
\end{equation}
We see that within the first approximation, both $\Sigma$ and
$\tilde{\Sigma}$ turn out to be momentum- and frequency-independent.
[It is easy to check that the next-order diagrams inevitably introduce momentum and
frequency dependence to self-energies and drastically change the structure of the theory.]
At this level of accuracy, the chemical potential equals to
$\mu = 2nU-n_0U$ according to the Hugenholtz-Pines relation. As mentioned above, it is
extremely convenient to use an effective chemical potential
\begin{equation}
\tm = \mu -2nU \;,
\label{eff_mu}
\end{equation}
as a thermodynamic variable to characterize properties of the WIBG. [Note that $\tm $
is negative.]
Within the same accuracy we can substitute $\tilde{\Sigma}$ with $-\tm\equiv |\tm| $ 
and simplify expressions for $G$ and $F$ to
\begin{equation}
G(P)\, =\, -{ {\rm i}\xi+\epsilon(k) + |\tm| \over \xi^2 +
E^2(k)}\,  , \label{G_dgr}
\end{equation}
\begin{equation}
F(P)\, =\,  {|\tm| \over \xi^2 + E^2(k)}\,  , \label{F_dgr}
\end{equation}
\begin{equation}
E^2(k)\, =\, \epsilon(k)[\epsilon(k)+ 2|\tm|] \, .
\label{E2}
\end{equation}

A note is in order here. Below we will calculate higher-order corrections to the chemical potential
which are necessary for the construction of the accurate equation of state,
see (\ref{mu_dgr}) and (\ref{rel_delta}). However, it is still possible
to use $\tm$  in the definitions of the propagators and the spectrum of elementary excitations
while keeping the accuracy of the entire scheme intact, see, subsection \ref{sec:Estimates} where
we study systematic errors involved in approximations. Here we mention briefly that the anomalous average
contribution to thermodynamic properties is always small, and thus further corrections to $F$
are negligible. The same is true for $G$ at low temperature. At temperatures
$T\gg n_0U$, on the other hand, one can ignore tiny modifications in the spectrum because
of an additional small parameter $n_0U/T$.

For the total density we have
\begin{eqnarray}
n\, =\, n_0-G(r=0,\tau=-0)\nonumber \\
=\, n_0+\sum_{\bf k}\, \left[ {\epsilon (k)- \tm \over 2E(k)}\, (1+2N_E)\,
-{1\over 2} \right] , \label{n_total}
\end{eqnarray}
where
\begin{equation}
N_\varepsilon \, =\, \left( {\rm e}^{\varepsilon/T} -1 \right)^{-1}
\label{N_E}
\end{equation}
is the Bose-Einstein occupation number for the mode with the energy $\varepsilon$. Formula (\ref{n_total}) includes
the leading and sub-leading terms.
To get an expression for the chemical potential with the same degree of accuracy,
we need to take into account higher-order diagrams and go beyond the leading-order expressions
for $\Sigma$ and $\tilde{\Sigma}$. This is because in the absence of interaction the
chemical potential is identically zero and thus the leading term is proportional to $U$.

\begin{figure}
\includegraphics[angle=-90, scale=0.25, bb=60 -400  500 200]{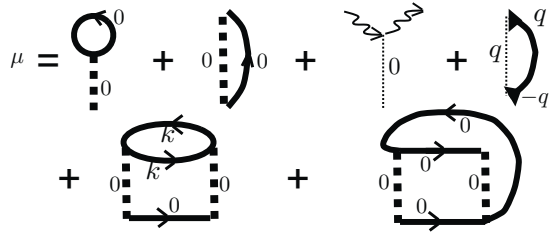}
\caption{Diagrams contributing to the chemical potential $\mu$ of a condensed gas, up to the second order. The last two diagrams
[that actually have  to be corrected according to (\ref{corrected})] are smaller than the previous ones and will be neglected; see the text for 
more detail.}
\label{fig:mu}
\end{figure}

There are three second-order diagrams contributing to $\Theta$. The first one is the anomalous Green's function convoluted with the bare interaction potential (see figure~\ref{fig:mu}.
The second one is the `sunrise' diagram with the proper correction (\ref{corrected}) for the
ladder structure and the third one is similar to the sunrise diagram but with propagators connecting different vertices (see the two diagrams in the second row in figure~\ref{fig:mu}). The latter two can be safely neglected away from the fluctuation regime because they involve an additional small parameter
$\gamma_0$ or $\gamma_T$, see Section~\ref{sec:Estimates} for the analysis and definitions of the fluctuation region and parameters $\gamma$ in different spatial dimensions.
It is worth noting that consistently taking into account contributions of the two neglected diagrams in the condensed regime would require 
simultaneously going to higher orders in the expansions for self-energies, figure~\ref{Fig:Sigma11}. The latter, however, is impossible without sacrificing
the attractive paradigm of independent quasiparticles with Bogoliubov dispersion. 

\begin{figure}
\includegraphics[angle=-90, scale=0.35, bb=170 -210  370 310]{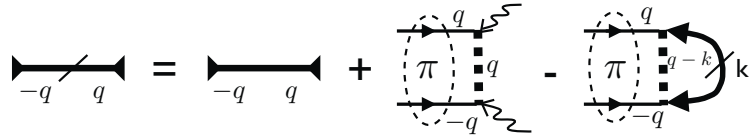}
\caption{Definition of $\Delta(\bf{q})$. }
\label{fig:Delta}
\end{figure}

Keeping the leading and the largest sub-leading diagrams for $\Theta$, we get
\begin{equation}
\mu \, =\, -2G(\tau=-0, r=0)U+n_0\, {\cal U}(0)-\sum_{\bf q}\,  {\cal U}({\bf q})F({\bf q},\tau=0) 
\, . \label{mu_dgr}
\end{equation}
We have no choice but to use the bare potential here because all ladder diagrams leading to the
pseudopotential vertex are already absorbed in the anomalous Green's function,
by construction of the latter. This feels unsatisfactory only at first glance since
simple formal manipulations allow one to express (\ref{mu_dgr}) in terms of $U$ alone.
Let us introduce an auxiliary  function $\Delta({\bf q})$ defined by the integral equation
(shown graphically in figure~\ref{fig:Delta})
\begin{equation}
\Delta({\bf q})= F({\bf q},\tau=0) + n_0\Pi({\bf q})U({\bf q})-\sum_{\bf k}\, \Pi({\bf q}-{\bf k}) U({\bf q}\! -\! {\bf k}) \Delta ({\bf k}) 
\, ,~ \label{delta}
\end{equation}
It can be used, in combination with the definition of the pseudopotential Eq.(\ref{eq:def_of_U}),
to transform the last two terms in (\ref{mu_dgr})
\begin{eqnarray}
n_0\, {\cal U}(0)-\sum_{\bf q} \, {\cal U}({\bf q})F({\bf q},\tau=0) \, =\, \nonumber \\
n_0U(0)-\sum_{\bf q}\,  U({\bf q})\Delta ({\bf q}) \, \approx\, (n_0-\Delta_0)U
\, , \label{rel_delta}
\end{eqnarray}
where
\begin{equation}
\Delta_0\, =\, \sum_{\bf q}\, \Delta ({\bf q})
\, . \label{delta_0}
\end{equation}
The last approximate equality in (\ref{rel_delta}) comes from $U({\bf q}) \approx U(0)$ and
the observation that $\Delta({\bf q})$ vanishes at momenta much smaller than $1/R_0$.


The smallness of the parameter $|U\Pi k^d|$ at momenta $k\ll 1/R_0$ allows one to
expand $\Delta$ [see  (\ref{delta})] in the series:
\begin{equation}
\Delta({\bf q})=\left[ F({\bf q},\tau=0) + n_0\Pi({\bf q})U({\bf q})\right] + \ldots
\; . \label{delta_series}
\end{equation}
For the effective chemical potential we thus get
\begin{equation}
\tm \, =\,-(n_0 + \Delta_0) U \; , \label{mu_delta}
\end{equation}
where within our order of accuracy we can take
\begin{equation}
\Delta_0\, =\, \sum_{\bf q} \, \left[ F({\bf q},\tau=0) + n_0U \Pi({\bf q})\right]  
\; . \label{delta_0_actual}
\end{equation}

We find it convenient to introduce a quantity with the dimension of density,
\begin{equation}
\n* \ = \ -\tm/U = n_0 + \Delta_0 \; .
\label{qc_def}
\end{equation}
 With this quantity
 the form of certain thermodynamic relations simplifies.
For example, using  (\ref{qc_def}) we can replace $n_0$ with $n_*$ in
(\ref{n_total}) and arrive at the result
\begin{equation}
n\, =\, \n*+ n' \, , \label{n_total_qc}
\end{equation}
\begin{equation}
n' = \sum_{\bf k}\, \left[  {\epsilon
(k) - E(k) \over 2E(k)} + \tm\Pi(k) +
{\epsilon (k) \over E(k)}\, N_E\right]  \, . \label{n_prime}
\end{equation}
This completes the self-consistent theory because we obtain a closed set of relations
which define $n=n(\mu,T)$ in the parametric form---given some $\tm$, or $n_*$,
one calculates $n$ from  (\ref{n_total_qc}) and (\ref{n_prime}) and then determines
$\mu$ from  (\ref{eff_mu}). The integral
in (\ref{n_prime}) is convergent not only in 3D, but also in 2D and 1D.
Hence, at this point the quantity $\eta$ can be set equal to zero.

We want to emphasize the fact that all specific expressions derived in this section feature
a two-parametric representational freedom, within the same order of accuracy.
First, it is possible to use $\tm$ with or without higher-order corrections in
the spectrum of elementary excitations and propagators. Second, there is a freedom of
choosing the function $\Pi(k)$.
For example, one could be tempted to absorb the first two terms in the integral
in (\ref{n_prime}) into the definition of $U$, such that $n'(T=0)$ gets identically equal to zero,
and $\n*(T=0)$ equals to the total number of particles. We, however, do not see any merit
in this protocol, because $U$ becomes dependent on $\tm$ and cannot be considered
as a fixed external parameter. Finally, this and analogous ``improvements'' do not make the theory
more/less accurate since the difference is of the same order as omitted diagrams. 

A remark is in order here concerning the two-dimensional case, where the value of $U$ cannot be defined irrespectively of the system density.
Even in this case, we can proceed with formally independent parameter  $\epsilon_0$ in (\ref{Pi_2D}) till we arrive
at the final answers along with the estimates of neglected higher-order terms (the latter being essentially $\epsilon_0$-dependent).
After that, the value of $\epsilon_0$ is selected in such a way that the order-of-magnitude values of neglected terms
are minimal.

\subsection{Pressure}
\label{subsec:pressure}

To derive relations for other thermodynamic properties, we start
with the pressure as a function of $T$ and $\tm$. Using
the general thermodynamic formula
\begin{equation}
n={\partial p(\mu,T) \over \partial \mu} \; , \label{rel1}
\end{equation}
and adopting---throughout this subsection---the convention that
$T$ is treated as a fixed constant, so that partial derivatives
with respect to either $\mu$ or $\tm$ can be replaced with
ordinary ones, we write
\begin{equation}
p\, =\, p_{\rm c} + \int_{0}^{\tm} n\, {{\rm d}\mu \over {\rm d} \tm}\, {\rm d}\tm \; ,
\label{p1}
\end{equation}
where
\begin{equation}
p_{\rm c}\, = \,n_{\rm c}^2 U -T \sum_{\bf k}\, \ln
\left[1-{\rm e}^{-\epsilon(k)/T} \right] \;  \label{p_c}
\end{equation}
is the value of pressure at the mean-field critical density
\begin{equation}
n_{\rm c}\, = \sum_{\bf k}\,  \left[{\rm
e}^{\epsilon(k)/T} -1  \right]^{-1} \;  . \label{n_c}
\end{equation}
Equations (\ref{eff_mu}) and (\ref{n_total_qc})  allow one to represent (\ref{p1}) as
\begin{equation}
p\, = \, p_{\rm c}-n_{\rm c}^2 U + \tm^2 /2U -2n'\tm + (n')^2 U  + \int_0^{\tm} n'\, {\rm d}\tm
\ \label{p2}
\end{equation}
(after  doing straightforward integrations by parts).
The integral in (\ref{p2}) is readily done by noticing that
\begin{equation}
{\epsilon (k) \over E(k)}\, =\, - {{\rm d} E(k) \over {\rm d}\tm}
\; , \label{obs_1}
\end{equation}
\begin{equation}
{\epsilon (k) N_E \over E(k)}\, = \, - T {{\rm d}  \over {\rm d}\tm} \,
\ln \left[1-{\rm e}^{-E(k)/T} \right]  \, , \label{obs_2}
\end{equation}
and thus
\begin{eqnarray}
n'\, =\, - {\frac 1 2} {{\rm d}\over {\rm d}\tm } \,    \sum_{\bf k}\,   [E(k) - \epsilon(k) + \tm -\tm^2 \Pi (k)] \nonumber \\
 - T  {{\rm d}\over {\rm d}\tm } \,   \sum_{\bf k}\,  \ln \left[1-{\rm e}^{-E(k)/T} \right]
\, , \label{obs_3}
\end{eqnarray}
where $\epsilon(k)$ makes the first term convergent. The result of the integration is
\begin{eqnarray}
\int_0^{\tm}n'\, {\rm d}\tm\, =\, \nonumber \\
 - \sum_{\bf k}\, 
\left\{  {E(k)-\epsilon(k)+\tm - \tm^2 \Pi(k) \over
2} + T\ln { 1-{\rm e}^{-E(k)/T} \over  1-{\rm e}^{-\epsilon(k)/T}  } \right\} \,  ,
\label{int}
\end{eqnarray}
and we finally get
\begin{eqnarray}
p\, = \,  \tm^2/2U - 2n'\tm + (n')^2 U  \nonumber \\
- {1\over 2}\sum_{\bf k}\, 
\left\{  E(k)-\epsilon(k) +\tm - \tm^2 \Pi (k)
+ 2T\ln \left[1-{\rm e}^{-E(k)/T} \right]
\right\}  \, . \label{p3}
\end{eqnarray}

\subsection{Entropy and energy}
\label{subsec:entropy}

Whenever  $n$, $\mu$, and $p$ are specified as functions of $(T,x)$,
where $x$ is a quantity of arbitrary nature, the expressions for entropy per unit volume, $s$,
and energy per unit volume, $\varepsilon$,
are readily found from the following two generic thermodynamic relations:
\begin{equation}
s=\left(\frac{\partial{p}}{\partial{T}}\right)_x-
n\left(\frac{\partial{\mu}}{\partial{T}}\right)_x\, , \label{s2}
\end{equation}
\begin{equation}
\varepsilon=\mu n+Ts-p\, .
\label{e1}
\end{equation}
With $x\equiv  \tm$ we thus find
\begin{eqnarray}
s=\sum_{\bf k}\, \left[{E(k)\over T} N_E -\ln\left(1-{\rm e}^{-E(k)/T}\right)
\right] \, . \label{s_cnd}
\end{eqnarray}
\begin{eqnarray}
\varepsilon= \frac{\tm^2}{2U} - n' \tm +n'^2U  \nonumber \\
+ {1\over 2}\sum_{\bf k}\, \left[ E(k)(2N_E+1)-\epsilon(k)+\tm -\tm^2 \Pi (k) \right] \, .
\label{e_cnd}
\end{eqnarray}

\subsection{Explicit integrations. $T=0$ case}
\label{subsec:practical}

It is useful to note that the following two integrals (we use $\tm\equiv -|\tm|$),
\begin{equation}
I_1^{(d)} (|\tm|)= {1\over 2}\sum_{\bf k}\, \left[ E(k)-\epsilon(k)-|\tm| -\tm^2 \Pi (k) \right] ,
\label{Int_1}
\end{equation}
\begin{equation}
I_2^{(d)} (|\tm|)= {1\over 2}\sum_{\bf k}\,  \left[  {\epsilon
(k)  \over E(k)} -2|\tm|\Pi(k) -1 \right]  = {\partial I_1^{(d)} \over \partial |\tm|} ,
\label{Int_2}
\end{equation}
entering relations for  thermodynamic quantities---see subsections \ref{subsec:basic}-\ref{subsec:entropy}---can be explicitly performed. [Here we assume that in 3D and 2D the kernel  $\Pi (k)$ is fixed
by expressions (\ref{Pi_3D}) and (\ref{Pi_2D}), respectively, and remind that in 1D it is zero.]
Explicitly doing the integrals is especially useful at $T=0$. In this case, all
other integrals in the expressions for thermodynamic functions nullify, and the answers reduce to algebraic relations. In the grand-canonical form, these relations read
\begin{equation}
|\tm| = \mu - 2UI_2^{(d)} (\mu) \qquad (T=0)\,  ,
\label{tm_groud}
\end{equation}
\begin{equation}
n = {\mu\over U} - I_2^{(d)} (\mu)\qquad  (T=0)\,  ,
\label{n_groud}
\end{equation}
\begin{equation}
p = {\mu^2\over 2U} - I_1^{(d)} (\mu)\qquad  (T=0)\,  .
\label{p_groud}
\end{equation}
Here we take into account that within our level of accuracy we can
replace $|\tm|\to \mu$ in the arguments of $I_1^{(d)} $ and $I_2^{(d)}$, since the integrals
are responsible for sub-leading corrections, while the sub-sub-leading terms are beyond our control.

Performing the integrations, one finds:
\begin{equation}
I_1^{(d=1)}(|\tm|) =-{2 \sqrt{m}\,  |\tm|^{3/2}\over 3 \pi}\, ,
\label{Int_1_1D}
\end{equation}

\begin{equation}
I_1^{(d=2)}(|\tm|) =  {m \tm^2 \over 8\pi } \left( \ln {|\tm|\over 2\epsilon_0}+{1\over 2} \right)  ,
\label{Int_1_2D}
\end{equation}

\begin{equation}
I_1^{(d=3)}(|\tm|) =  {8m^{3/2}|\tm|^{5/2}\over 15\pi^2}\, ,
\label{Int_1_3D}
\end{equation}
and differentiating with respect to $|\tm|$ we find

\begin{equation}
I_2^{(d=1)}(|\tm|) = -{\sqrt{m |\tm|}\over \pi}\, ,
\label{Int_2_1D}
\end{equation}

\begin{equation}
I_2^{(d=2)}(|\tm|) =  {m |\tm|\over 4\pi } \left( \ln {|\tm|\over 2\epsilon_0}+1\right)  ,
\label{Int_2_2D}
\end{equation}

\begin{equation}
I_2^{(d=3)} (|\tm|) =  {4m^{3/2}|\tm|^{3/2} \over 3\pi^2}\, .
\label{Int_2_3D}
\end{equation}

In 2D, we have a freedom of fine-tuning $\epsilon_0$ to simplify the form of the answer. A natural choice, especially convenient for the $T=0$ limit, is to set
\begin{equation}
\epsilon_0 = ({\rm e}/2)|\tm| \, ,
\label{eps_best}
\end{equation}
in which case we have
\begin{equation}
I_2^{(d=2)}\equiv 0\, , \qquad  I_1^{(d=2)} =  -{m \tm^2 \over 16 \pi } \, .
\label{Int_1_2_2D}
\end{equation}
At $T=0$ this translates into very compact grand-canonical expressions \cite{Mora}
\begin{equation}
|\tm| =\mu \, , \qquad  n={\mu\over U}  \qquad (d=2,\quad T=0) \, ,
\label{2D_ground_1}
\end{equation}
\begin{equation}
p={\mu^2\over 2U}\left( 1+ {mU\over 8\pi }\right)  \qquad  (d=2, \quad T=0)\,  ,
\label{2D_ground_2}
\end{equation}
with
\begin{eqnarray}
U = {4\pi/m \over \ln (4/a^2m\mu) -2\gamma - 1}  ,  \quad  (d=2, \quad T\lesssim T_{\rm c}).
\label{U_fine}
\end{eqnarray}
It is seen that the simplicity of the form of (\ref{2D_ground_1}) comes at the expense of more sophisticated (fine-tuned) form of the effective interaction (\ref{U_fine}). If, instead,
one would opt to simplify the form for the effective interaction, getting rid of sub-logarithmic terms: $U=(4\pi /m)/\ln (1/a^2m\mu)$, then (\ref{2D_ground_1}) would
acquire the generic form (\ref{tm_groud})-(\ref{n_groud}). Needless to note that this does not change the {\it sum} of leading and sub-leading terms in the equations of state since, by construction, the result cannot depend on the specific choice of $\epsilon_0$.

\subsection{Superfluid density}

\label{subs:SD}

The standard way of calculating the superfluid density within the quasi-particle picture
is based on Landau's formula for the normal component density, $n_{\rm n}$, and the relation
$n_{\rm s}=n-n_{\rm n}$. Here we employ a more general approach based on the current induced
by the gradient of the condensate wave-function.
By definition, the superfluid density is the linear-response
coefficient relating the persistent-current density to the gradient of the phase, $\varphi$, of the
(complex) order parameter field:
\begin{equation}
{\bf j}=\frac{n_{\rm s}}{m}\nabla \varphi = n_{\rm s} \frac{{\bf k_0}}{m} \, .
\label{ns1}
\end{equation}
In the last equality we assumed that $\varphi = {\bf k}_0\cdot{\bf r}$. On the other hand,
the average current can be calculated from the microscopic operator expression
in terms of the condensate density and the Green's function
\begin{equation}
{\bf j}=-\frac{\rm i}{2m}\:\left \langle [ \hat{\psi}^{\dag}  \nabla \hat{\psi} -{\rm H.c.} ] \right\rangle =
\frac{{\bf k}_0}{m} n_0 + \sum_{\bf k}\, \frac{{\bf k}}{m} \: N({\bf k}_0,{\bf k}) ,
\label{ns2}
\end{equation}
where
\begin{equation}
N({\bf k}, {\bf k}_0)=G(\tau=-0, {\bf k})\vert_{{\bf k}_0}
\; .
\label{Nkk}
\end{equation}
In the limit of $k_0\to 0$ we obtain the required relation
\begin{equation}
n_{\rm s}= n_0 + \lim_{k_0\to 0}\sum_{\bf k}\, \frac{{\bf k}\cdot {\bf k}_0}{k_0^2} \: N({\bf k}_0,{\bf k}) \;.
\label{ns3}
\end{equation}

At first glance equation (\ref{ns3}) does not resemble Landau's formula at all, since
the first term is given by the condensate density, not $n$. However, the structure of the
normal average contribution is such that it has a part which completes $n_0$ to the total
density and a part which equals (with minus sign) to the normal density component.
To derive this result, we start with the expression for the Green's function
\begin{equation}
G(\xi , {\bf k})\vert_{{\bf k}_0} = \frac{ {\rm i} \xi +\epsilon({\bf k}\! -\! 2{\bf k}_0) - \tm }{\tm^2+[ {\rm i} \xi +
\epsilon({\bf k}\! -\!  2{\bf k}_0) - \tm] [ {\rm i} \xi -\epsilon({\bf k})+\tm ] } \, ,
\label{Nkk1}
\end{equation}
and solve for the roots of the denominator in (\ref{Nkk1})
\begin{eqnarray}
R_{1,2}=\frac{\epsilon(k)-\epsilon({\bf k}\! -\!  2{\bf k}_0)}{2} \pm  \nonumber \\
\sqrt{\epsilon({\bf k})\epsilon({\bf k}\! -\!  2{\bf k}_0) -[\epsilon({\bf k})\! +\! \epsilon({\bf k}\! -\!  2{\bf k}_0)]\tm+\left[ \frac{\epsilon({\bf k})-\epsilon({\bf k}\! -\!  2{\bf k}_0)}{2}\right]^2}
\label{Nkk2}
\end{eqnarray}
to rewrite the Green's function identically as
\begin{equation}
G=\frac{\alpha}{i\xi-R_1}+ \frac{1-\alpha}{i\xi-R_2}\, ,
\nonumber
\end{equation}
\begin{equation}
\alpha=\frac{\epsilon({\bf k}- 2{\bf k}_0) -\tm - R_1}{R_1-R_2} \, .
\label{Nkk3}
\end{equation}
Now we express frequency sums through the Bose functions to arrive at
\begin{equation}
N({\bf k})\vert_{{\bf k}_0}= \alpha N(R_1)+(\alpha -1)(1+N(-R_2))   \, .
\label{Nkk4}
\end{equation}

The rest of the calculation is  straightforward. In view of the limit to be taken
in (\ref{ns3}) it is sufficient to keep only terms linear in
$[\epsilon({\bf k})-\epsilon({\bf k}\! -\! 2{\bf k}_0)] \approx 2{\bf k}\cdot {\bf k}_0/m$ in $R_{1,2}$, 
$\alpha$, and $N({\bf k})\vert_{{\bf k}_0}$. We omit the algebra of doing the expansion which
leads to the result
\begin{equation}
n_{\rm s}=n_0+\sum_{{\bf k}}  \frac{k^2}{dm} \: \frac{\tm^2}{E^2(k)} \:
\left[ \frac{1+2N_E}{2E(k)}-\frac{\partial N }{ \partial E(k)} \right] .
\label{ns4}
\end{equation}
Finally, using (\ref{n_total}) to exclude $n_0$
and integrating by parts to simplify the expression we arrive at Landau's formula
\begin{equation}
n_{\rm s}=n+{1\over d}\sum_{\bf k}\, \frac{k^2}{m} \frac{\partial N }{ \partial E}\,  .
\label{nslandau}
\end{equation}

\section{Expansion parameters. Estimates for higher-order terms}
\label{sec:Estimates}

Let us analyze the structure of small parameters which control
the applicability of the diagrammatic technique presented above.
As we will see {\it a posteriori}, it is sufficient to consider two
characteristic limits: (i) the $T=0$ case and (ii)
the finite-$T$ contributions of diagrams with zero frequencies (i.e., classical-field contributions).
The analysis is based on dimensional estimates of the diagrams, the fact that an extra interaction
vertex increases the total number of propagators by two, and that the largest contributions
come from small momenta (and frequencies), where the normal and anomalous propagators behave as
\begin{equation}
|G|\, \approx \, |F|\, \approx \,  {|\tm|  \over \xi^2 + c^2k^2} \, , \qquad c^2=|\tm | /m \, ,
\label{G_F_est}
\end{equation}
and we do not need to distinguish between them. We will further assume that there is an infra-red
momentum cutoff $k_1 \ll k_0 $ where
\begin{equation}
k_0 = \sqrt{m n U}
\label{k_0def}
\end{equation}
is directly related to the inverse healing length.
The symmetry-breaking field $\eta$ also changes the behaviour of propagators
at $k\ll k_0$, but for our purposes we do not need the explicit form of
propagators at finite $\eta$ because we will keep $k_1$ large enough to neglect effects
originating from finite $\eta$.

At $T<T_{\rm c}$ there are two kinds of generic vertices in higher-order diagrams, namely
(i) full vertices, where four propagators meet, and (ii) vertices with one condensate line
(vertices with two and three condensate lines are all accounted for by the lowest-order diagrams
for the self-energies and can not be part of the diagrammatic expansion).
A naive definition of the diagram order as the total number of vertices proves inconvenient, since contributions of condensate vertices turn out to be larger than contributions of full vertices. Below we will see that the difference is exactly compensated by the fact that condensate vertices (generically) come in pairs.  This suggests to define the diagram order as the sum of the total number of full vertices and half the total number of condensate vertices (plus $1/2$ for anomalous diagrams).

\subsection{$T=0$ case}

At $T=0$, the  structure for the dimensionless factor associated with adding an extra full vertex to a diagram is
\begin{equation}
\gamma^{\rm (full)}_0\, \sim \, U\int \delta(\Delta k)\delta(\Delta \xi)  \left[|G|\,  {\rm d}^dk\,  {\rm d}\xi \right]^2\, ,
\label{full_0_rel1}
\end{equation}
with $\delta(\Delta k)$ and $\delta(\Delta \xi)$ representing the $\delta$-functions taking care of momentum and frequency conservation laws. 
The dimensional estimate, following from (\ref{full_0_rel1}) and  (\ref{G_F_est}), reads
\begin{equation}
\gamma^{\rm (full)}_0\, \sim \, \frac{n^{1/2}(mU)^{3/2}}{k_1^{3-d}}\, .
\label{full_0_rel2}
\end{equation}

The  structure for the dimensionless factor associated with adding an extra pair of
condensate vertices is
\begin{equation}
\gamma^{\rm (cnd)}_0\, \sim \, U^2 n \int [\delta(\Delta k)\delta(\Delta \xi)]^2  \left[|G|\,  {\rm d}^dk\,  {\rm d}\xi \right]^3\, ,
\label{cnd_0_rel1}
\end{equation}
and from the corresponding dimensional estimate one can see that
\begin{equation}
\gamma^{\rm (cnd)}_0\, \sim \, \gamma^{\rm (full)}_0 \  {mnU \over k_1^2} \, .
\label{cnd_0_rel2}
\end{equation}
Equations (\ref{full_0_rel2}) and (\ref{cnd_0_rel2}) clearly show the infrared problem of the theory in dimensions $d\leq 3$ \cite{Nepom78,Popov79,Nepom83} 
(as usual, a zero power of the momentum  cutoff  should be understood as a logarithm), and emphasize the usefulness of the cutoff-enforcing field $\eta$. 
For the diagrammatic expansion to be consistent, we need $k_1$ to be large enough in order to guarantee $\gamma_0 \ll 1$. 
On the other hand, we do not want the field $\eta$ to significantly affect the physics of the system, and thus need $k_1 \ll  k_0$.
Hence, the smallest possible expansion parameter one can afford without distorting the physics corresponds
to $k_1 \le k_0$, in which case
\begin{equation}
\gamma^{\rm (full)}_0\, \sim \, \gamma^{\rm (cnd)}_0\, \sim\, \gamma _0 \; = \;\left\{ {\begin{array}{*{20}c}
  {\sqrt {na^3 } \qquad (d = 3),}  \\
  {mU\qquad  (d = 2),}  \\
  {\sqrt {mU/n} \qquad (d = 1) .} \\
\end{array}} \right.
\label{gamma_0_rel3}
\end{equation}
Here we took into account that $|\tm (T=0)| \sim nU$, and expressed $U$ in terms of $a$ in 3D.
It is clear that $\gamma_0$, equation  (\ref{gamma_0_rel3}), is the actual expansion parameter for all local thermodynamic quantities. 
These quantities, by referring to large length scales at $T=0$,
should mainly have contributions from wavevectors $\lesssim k_0$. Technically, it means that
infrared divergencies of individual diagrams have to cancel each other
in the final answers and the resulting integrals become convergent at the wavevectors  $k\sim k_0$, leading
to a well defined expansion in powers of $\gamma_0$ (\ref{gamma_0_rel3}).
[Since dimensional estimates do not distinguish between pure powers and powers with
logarithmic pre-factors, we extend the meaning of the term `in powers' to the powers with
logarithmic pre-factors.]

Note that all our relations between the thermodynamic quantities that do not vanish in the $T=0$ limit  (specifically, $n$, $\mu$, $p$, $\varepsilon$, and $n_0$),
are accurate up to the first-order corrections in $\gamma_0$, so that their systematic errors are of the order of $\gamma_0^2$ (with possible logarithmic pre-factors).
The relation for the entropy is accurate only up to the leading term, which actually has the same  quasi-particle origin as the {\it sub-leading} 
terms in all non-vanishing at $T\to 0$ quantities. The same is true for the heat capacity that behaves similarly to entropy at $T\to 0$.

\subsection{Finite-$T$ zero-frequency contributions}
\label{subsec:finite_T}

The  dimensionless factor associated with adding an extra full vertex to a diagram consisting of zero-frequency propagators scales as
\begin{equation}
\gamma^{\rm (full)}_T\, \sim \, (U/T) \int \delta(\Delta k)  \left[ |G|\,T\,   {\rm d}^dk\,  \right]^2\, ,
\label{full_T_rel1}
\end{equation}
yielding the dimensional estimate
\begin{equation}
\gamma^{\rm (full)}_T\, \sim \, \frac{m^2TU}{k_1^{4-d}}\, .
\label{full_T_rel2}
\end{equation}
For the pair of condensate vertices we have
\begin{equation}
\gamma^{\rm (cnd)}_T\, \sim \, (U/T)^2n_0 \,  \int [\delta(\Delta k)]^2  \left[ |G|\,T\,   {\rm d}^dk\,  \right]^3\, .
\label{cnd_T_rel1}
\end{equation}
Similarly to the $T=0$ case (assuming that $|\tm| \sim n_0U$, we get
\begin{equation}
\gamma^{\rm (cnd)}_T\, \sim \, \gamma^{\rm (full)}_T \  {mn_0 U \over k_1^2} \, .
\label{cnd_T_rel2}
\end{equation}

In complete analogy with the $T=0$ case, the largest possible $k_1$ (enforced by $\eta$) cannot exceed $\sqrt{mn_0U}$. 
Assuming that $k_1 \le \sqrt{mn_0U}$ results in
\begin{equation}
\gamma^{\rm (full)}_T\, \sim \, \gamma^{\rm (cnd)}_T\, \sim \, \gamma_T\; \sim \;  \gamma_0 \, {T\over nU} \, \left( {n\over n_0} \right)^{2-d/2}\, .
\label{gamma_T_rel3}
\end{equation}
This expression shows that at $T\ll nU$ the parameter $\gamma_0$ dominates over $\gamma_T$, and thus this temperature regime is equivalent to $T=0$.
In the crossover regime, $T\sim nU$, both $\gamma_0$ and $\gamma_T$ are of the same order.
In the regime  $T\gg nU$, the classical-field parameter $\gamma_T$ dominates, and becomes of order unity when $n_0$ gets sufficiently small. The latter situation corresponds
to the fluctuation region, where the perturbative theory becomes inadequate, since it fails to properly describe the non-linear long-wave fluctuations of the classical component of the quantum field.  Before hitting the region $\gamma_T\sim 1$, one has to switch to the description in terms of scaling functions \cite{Tc3D,Tc2D} which are universal for all $U(1)$ weakly-interacting theories, see section \ref{sec:Numerics}.

At $T\gg nU$,  all our relations for the thermodynamic quantities are accurate up to the first-order corrections in $\gamma_T$, 
so that their systematic errors are of the order of $\gamma_T^2$ (with possible logarithmic pre-factors). 
Note that at  $T\gg nU$ the leading terms for all the thermodynamic quantities, except for the chemical potential 
(and with a reservation for the density in 1D where the corresponding condition is  $T\gg nU/\gamma_0$), are
the same as for the ideal gas.

To summarize our analysis, we present below an order-of-magnitude estimate of systematic uncertainties
due to omitted higher-order terms for each thermodynamic quantity.
It is convenient to start with $\mu$ and recall that the omitted term in (\ref{mu_dgr})
comes from the `sunrise' diagram for $\Theta$. A dimensional analysis of this diagram gives
\begin{equation}
\Delta \mu /\tm  \, \sim\, \max[\gamma_0^2, \gamma_T^2]\; . \label{Delta_mu}
\end{equation}
which transforms into a similar estimate for the total and superfluid density
\begin{equation}
\Delta n / \n* \, \sim\,  \max[\gamma_0^2, \gamma_T^2]\; . \label{Delta_n}
\end{equation}
\begin{equation}
\Delta n_{\rm s}\, \sim\, \Delta n\; . \label{Delta_n_qc}
\end{equation}
This can be seen from the dimensional analysis of the second-order diagrams contributing to the self-energies.
Relation (\ref{p1}) for the pressure implies
\begin{equation}
\Delta p\, \sim\, n \tm  \max[\gamma_0^2, \gamma_T^2]\; . \label{Delta_p}
\end{equation}
In the case of entropy, it is convenient to start with the regime $T\gtrsim nU$ when---using  (\ref{s2}) to relate the uncertainty in $s$ to $\Delta p$, $\Delta n$, and $\Delta \mu$---we find
\begin{equation}
\Delta s\, \sim\, {\tm \over T}\, n\,   \gamma_T^2\qquad (T\gtrsim nU) \; . \label{Delta_s_highT}
\end{equation}
We see that at $T\sim nU$, in contrast to the behaviour of other thermodynamic functions, the uncertainty in $s$ scales only as the first power of $\gamma_0$ (simply because $s$ itself gets of order
$\sim \gamma_0$).
This scaling persists down to $T\to 0$:
\begin{equation}
\Delta s /s \, \sim\, \gamma_0 \qquad (T\lesssim nU) \; , \label{Delta_s_lowT}
\end{equation}
because, at $T\ll nU$,  the thermodynamics of the system corresponds to the generic low-temperature behaviour of superfluids, where the leading temperature-dependent contributions
are due to phonons.  Hence, the first-order correction to the sound velocity, which is of the order of $\gamma_0$ as is  seen from the expression for the
energy, translates into the order-$\gamma_0$ correction to the entropy. At $T\ll nU$, when thermodynamics is exhausted by dilute non-interacting phonons, the correction to the sound velocity can be found directly from the zero-temperature compressibility, and the accuracy of the expression for $s$ (and other thermodynamic quantities) can be improved. At $T\sim nU$, however, the order-$\gamma_0$ correction to the quasiparticle dispersion law goes beyond the Bogoliubov  ansatz \cite{Fedichev}. Note also that improving the value of the sound velocity  in the $T \ll nU$, while rendering the answers for thermodynamic quantities more accurate,  is inconsistent with retaining the Bogoliubov form of the spectrum for {\it all} the momenta.

Finally,  equation (\ref{e_cnd}) in combination with the above results yields the estimate for the uncertainty of energy:
\begin{equation}
\Delta \varepsilon\, \sim\, \Delta p\, \sim\, n \tm  \max[\gamma_0^2, \gamma_T^2] \; . \label{Delta_e}
\end{equation}

\subsection{Fluctuational contributions}
\label{subsec:fluct_contr}

In all spatial dimensions, there is an  interval (in terms of temperature, if density is kept fixed, or in terms of chemical potential at fixed temperature, etc.) where $\gamma_T$ becomes of order unity, and the systematic perturbative description breaks down. In 3D and 2D this happens in the vicinity of the superfluid phase transition point. In 1D there is no superfluidity in the strict sense of the word, and no phase transition occurs, but the picture remains  very similar to that of 2D and 3D case for a weakly interacting gas, as  explained in the next subsection. A detailed discussion of the fluctuation region, including accurate expressions for the critical points,  will be presented later, in section \ref{sec:fluct}.  Meanwhile, here we want to utilize the fact that the results (\ref{Delta_mu})-(\ref{Delta_e}) allow one to make order-of-magnitude estimates of the non-perturbative fluctuation contributions to the thermodynamic functions. These estimates are quite important as they show the  degree to which the perturbative results of the previous sections are inaccurate in the fluctuation region. As we will see, for some quantities the fluctuation corrections are smaller than the leading ideal-gas contributions.

By continuity, the order-of-magnitude estimates for fluctuation contributions can be obtained from (\ref{Delta_mu})-(\ref{Delta_e}) by simply setting $\gamma_T \sim 1$, while for the quantities themselves we can use their mean-field critical expressions.
This way we arrive at the following results
\begin{equation}
{(\Delta \mu)_{\rm fluct} \over \mu} \, \sim {(\Delta n)_{\rm fluct} \over n} \, \sim \, \left\{ {\begin{array}{*{20}c}
{ \gamma_0^{2/3} \qquad (d = 3)\, , } \\
  { \ln^{-1} (1/\gamma_0) \qquad (d = 2)\, , }  \\
    {1 \qquad (d = 1)\, . }  \\
\end{array}} \right.
\label{n_tilde_mu_tilde}
\end{equation}
\begin{eqnarray}
{(\Delta p)_{\rm fluct}\over p}\, \sim\,   {(\Delta \varepsilon)_{\rm fluct}\over \varepsilon}\, \sim\,  {(\Delta s)_{\rm fluct} \over s}\, \sim \nonumber \\
\sim \, \left\{ {\begin{array}{*{20}c}
{  \gamma_0^{4/3} \qquad (d = 3)\, , } \\
  { \gamma_0 \ln ( 1/\gamma_0) \qquad (d = 2)\, , }  \\
    {\gamma_0^{1/2}\qquad (d = 1)\, . }  \\
\end{array}} \right.
\label{p_e_s_fluct}
\end{eqnarray}

For the superfluid and condensate densities we get an obvious answer that in both cases the fluctuation
contributions are on the order of 100\%.

\subsection{Specifics of 1D system}
\label{subsec:specifics_1D}

As long as a 1D system is weakly interacting,  i.e. $\gamma_T \ll 1$, the notion of the order parameter field with well defined amplitude and the two-component (normal + superfluid) description remain physically meaningful. Since superfluidity is a topological phenomenon,
it can be destroyed  only by topological defects---phase slips. At  $\gamma_T \ll 1$ the
phase slips are rare events which do not contribute significantly to the local
thermodynamic quantities; in correlation functions, phase slips show up only  at length scales much larger than $1/k_0$, where their effect can be described at the hydrodynamic level.

When the temperature reaches the characteristic scale of
\begin{equation}
T_{\rm fluct}^{(1D)}\, \sim \, \gamma_0^{-1}  nU\,  \sim\, \gamma_0 {n^2\over m}  \,  , \label{fluct_1D}
\end{equation}
the parameter $\gamma_T$ becomes of order unity.
Apart from the lack of a genuine phase transition, the physics in the temperature range $T\sim T_{\rm fluct}^{(1D)}$ is close to that of the fluctuation region in 2D and 3D. Only the long-wave classical-field subsystem of the original quantum-field experiences strong non-linear fluctuations.
This leads to non-perturbative contributions to the system thermodynamics which are universal for all weakly interacting one-dimensional U(1) systems, and can be described by universal scaling functions in direct analogy with the fluctuation regions in 2D and 3D \cite{Tc3D,Tc2D}. At $T\gg T_{\rm fluct}^{(1D)}$, the low-momenta part of the classical-field component gets depleted, so that the non-linearity of
interactions becomes weak and accurately accounted for within the normal-gas mean-field picture.

\section{Off-diagonal correlations}
\label{sec:off}

Beliaev's diagrammatic technique allows one to calculate correlation functions up to distances much larger than the correlation radius $r_0\sim 1/k_0$.
However, addressing the asymptotic long-range behaviour of off-diagonal correlation functions requires special techniques properly accounting for long-wave fluctuations of the Goldstone mode, i.e. the phase of the superfluid order parameter. Since the diagrammatic and hydrodynamic description overlap,
one can straightforwardly extend the diagrammatic calculation of off-diagonal correlators obtained
up to large enough distances, $r \gg r_0$ to arbitrarily large $r$'s. We proceed in the spirit of Popov's hydrodynamic approach \cite{Popov79}, but with a significant simplification that the hydrodynamic treatment takes place on top of Beliaev's techniques, and thus does not require any additional modifications. The simplification is possible due to the fact that
at large distances only phase fluctuations are important and their effect can be factored out as
\begin{equation}
\psi \, =\,\tilde{ \psi} \, {\rm e}^{ {\rm i} \Phi} \,  , \label{factor_out}
\end{equation}
where, under coarse-graining up to momentum $k_1$,
\begin{equation}
\Phi({\bf r}, \tau)\, =\, \sum_{k<k_1}  {\rm e}^{{\rm i} {\bf k}{\bf r}} \Phi({\bf k}, \tau)  \,    \label{Phi}
\end{equation}
is the long-wave part of the phase field. As before, the momentum  cutoff $k_1\ll k_0$
should be small enough to guarantee (i) the statistical independence
of the fluctuations of $\Phi$ from the fluctuations of $\tilde{ \psi}$, and (ii)
the Gaussian character of phase fluctuations in the hydrodynamic regime.
(A particular value of $k_1$ is not important and does not appear in final expressions.)

In view of (i) and (ii), we have for an $m$-particle correlator,
\begin{equation}
{\cal K}\, =\, \langle \, \psi_1\,  \psi_2^* \, \psi_3\,  \psi_4^*\, \cdots \, \psi_{2m-1}\,  \psi_{2m}^*\, \rangle \, ,   \label{K_cor_0}
\end{equation}
where the subscripts label the space-time variables. Using (\ref{Phi}) it can be written as
\begin{equation}
{\cal K}\, =\, \tilde{\cal K} \, {\rm e}^{-\Lambda}\, ,   \label{K_cor_1}
\end{equation}
with
\begin{equation}
\Lambda\, =\, {1\over 2} \Big\langle \, \Big( \sum_{j=1}^{2m} (-1)^j\Phi_j \Big)^2 \, \Big\rangle\, ,  \label{Lambda}
\end{equation}
where $\tilde{\cal K}$ is obtained from ${\cal K} $ by substituting $\psi_j \to \tilde{\psi}_j $, for all $j$'s.

At distances $|{\bf r}_ j -{\bf r}_ s | \gg 1/k_0$ the correlator $\tilde{\cal K}$ remains constant, and the
remaining dependence on distance is exclusively due to the fluctuations of $\Phi$.
[The correlator $\tilde{\cal K}$ does depend on the coarse-graining momentum $k_1$
and this dependence is crucial for compensating the $k_1$-dependence of the function $\Lambda$; in fact
its origin is directly related to $\Lambda$ when coarse-graining is stopped at larger momenta
and then taken further to $k_1$.]
The independence of  $\tilde{\cal K}$ on coordinates and times in the asymptotic limit
immediately leads to the relation
\begin{equation}
{\cal K}(\vec{X})\, =\, {\cal K}(\vec{X}') \, {\rm e}^{\Lambda(\vec{X}') -\Lambda(\vec{X})}\,   \label{K_K}
\end{equation}
between correlators at different sets of variables, $\vec{X}$ and $\vec{X}'$, provided both are in the asymptotic region. Now if  $\vec{X}'$ is within reach of diagrammatic expansions,
then (\ref{K_K}) allows one to ``extrapolate''  ${\cal K}(\vec{X}') $ to  ${\cal K}(\vec{X}) $ at {\it arbitrary} $\vec{X}$ in the asymptotic domain. By the structure of (\ref{K_K}), the difference $\Lambda(\vec{X}') -\Lambda(\vec{X})$ is independent of $k_1$, as opposed to individual $\Lambda$'s.
Here we assumed for simplicity that all variables in $\vec{X}$ and $\vec{X}'$ are large enough.
The same analysis is readily adapted for cases when only some of the
coordinates in the $m$-particle correlation function are in the asymptotic regime; only these
coordinates have to be mentioned in (\ref{K_K}).

The Gaussian character of the field $\Phi$ implies, see (\ref{Lambda}), that
\begin{equation}
\Lambda(\vec{X}) -\Lambda(\vec{X}')\, =\, \sum_{s<j}\, (-1)^{s+j}\,  \Xi({\bf r}_{sj}, \tau_{sj}; {\bf r}_{sj}', \tau_{sj}') \, ,
\label{Lambda_Lambda}
\end{equation}
where ${\bf r}_{sj}={\bf r}_{s}-{\bf r}_{j}$, $\tau_{sj}=\tau_{s}-\tau_{j}$ (the same for primed variables), and
\begin{eqnarray}
\Xi({\bf r}, \tau; {\bf r}', \tau') \, =\, \langle\,  \Phi({\bf r},\tau)\Phi({\bf 0},0)- \Phi({\bf r}',\tau')\Phi({\bf 0},0)\, \rangle \nonumber \\
=\, T \sum_{\xi, {\bf k}}\, \left[ {\rm e}^{{\rm i}({\bf k}{\bf r}-\xi \tau)} -{\rm e}^{{\rm i}({\bf k}{\bf r}'-\xi \tau')}\right] \left\langle \, |\Phi({\bf k},\xi)|^2\right\rangle\, .
\label{Xi_def}
\end{eqnarray}
We see that for the extrapolation of correlation functions to arbitrarily large distances
one needs to know only  $\Xi({\bf r}, \tau; {\bf r}', \tau')$ which is defined through the
average $\left\langle \, |\Phi({\bf k},\xi)|^2\right\rangle$. The latter is readily found
from Popov's hydrodynamic action
\begin{equation}
S\, =\, {T\over 2} \sum_{\xi, {\bf k}} \,  [(n_{\rm s}/m)k^2+\kappa \xi^2] |\Phi({\bf k},\xi)|^2 \, ,
\label{hydro}
\end{equation}
leading to
\begin{equation}
\left\langle \, |\Phi({\bf k},\xi)|^2\right\rangle\, =\, \left[ (n_{\rm s}/m)k^2+\kappa \xi^2 \right]^{-1}  .
\label{phase_fluct}
\end{equation}

In fact, non-zero frequencies (accounting for the quantization of the fluctuations of the phase) are relevant only at $T\ll nU$, meaning that
the parameter $\kappa$ in (\ref{phase_fluct}) can be taken in its $T=0$ limit,
\begin{equation}
\kappa\,  =\,  \left( {{\rm d}\mu \over {\rm d} n}\right)^{-1}_{T=0}\,  \approx U^{-1}\, .
\label{kappa}
\end{equation}
At $T\gtrsim nU$, only the $\xi=0$ term should be left in (\ref{Xi_def}).

Let us consider now the single-particle density matrix
\begin{equation}
\rho({\bf r})\, =\, \langle \psi^*({\bf r})\psi({\bf 0}) \rangle\, .
\label{spdm}
\end{equation}
Taking into account that $\rho({\bf r})\equiv n + G(r=0, \tau=-0) - G(r, \tau=-0)$, from the diagrammatic expressions for $G$ we have (at small and intermediate $r$'s)
\begin{equation}
\rho({\bf r}) = n- \sum_{\bf k} \,  \Big( 1- {\rm e}^{ {\rm i} {\bf k}{\bf r}}\Big) \Big[ {\epsilon (k) - \tm \over 2E(k)}\ (1+2N_E)
-{1\over 2} \Big]   . \label{spdm_2}
\end{equation}
We can use this expression in any dimension at distances significantly exceeding $1/k_0$.
In 2D at finite temperature and in 1D at any temperature, the expression (\ref{spdm_2})
ultimately becomes inaccurate, and we have to rely on the above-described procedure of extrapolation.
In the asymptotic regime we have
\begin{eqnarray}
\rho({r})\, =\, \rho({ r}')\, {\rm e}^{-\Xi(r,r')}\; ,\label{spdm_extr}
\end{eqnarray}
where, at $T\ll nU$,
\begin{eqnarray}
\Xi(r,r') \, =\, {1\over \kappa} \sum_{\bf k} \, \Big[ {\rm e}^{ {\rm i} {\bf k}{\bf r}'} -  {\rm e}^{ {\rm i} {\bf k}{\bf r}} \Big] {1+2N_{E=ck}\over 2ck}\; ,~~~\label{Xi_low}
\end{eqnarray}
with $c=\sqrt{n_{\rm s}/m\kappa}$ the sound velocity at $T=0$ (here $n_{\rm s}=n$), while at $T\gtrsim nU$:
\begin{eqnarray}
\Xi(r,r') \, =\, {mT\over n_{\rm s}} \sum_{\bf k} \, \Big[ {\rm e}^{ {\rm i} {\bf k}{\bf r}'} -  {\rm e}^{ {\rm i} {\bf k}{\bf r}} \Big] {1 \over k^2}\; .~~~\label{Xi_classical}
\end{eqnarray}

Within the relevant orders, equation (\ref{spdm_2}) corresponds to the expansion of the exponential in (\ref{spdm_extr}) in powers of $\Xi$, up to the term $\propto \Xi$ included.  This immediately suggests that within the same accuracy one can exponentiate $\Xi$-terms in (\ref{spdm_2}) to extend its domain of applicability to much larger distances. The other advantage of proceeding this way
is having a physical definition of the quasi-condensate density as the amplitude of the
order parameter field in the long-wave limit \cite{Svistunov}. The equations that follow have the same
accuracy as (\ref{spdm_2}) though they do contain artificial higher-order terms which arise from factorizing the correlation function. More specifically, we single out terms in (\ref{spdm_2})
which on large distances reproduce $\Xi$, and then exponentiate them:
\begin{equation}
\rho({\bf r}) = [n - d({\bf r})- n \Lambda ({\bf r})] \ \longrightarrow  \
\tilde{\rho}({\bf r})\,  {\rm e}^{-\Lambda ({\bf r}) } \;,
\label{spdm_3}
\end{equation}
where $\tilde{\rho}({\bf r}) \ = \ n- d({\bf r}) $, and
\begin{equation}
d({\bf r}) =  \sum_{\bf k} \, \Big( 1- {\rm e}^{ {\rm i}{\bf k}{\bf r}} \Big)  \frac{\epsilon }{E^2}
\Big[ \frac{E-\epsilon+\tm }{2} +(E - \tm )N_E \Big] \;,
\label{ntilde}
\end{equation}
\begin{equation}
\Lambda ({\bf r}) =  - \frac{\tm }{n}\sum_{\bf k} \, \Big( 1- {\rm e}^{ {\rm i}{\bf k}{\bf r}}\Big)
\frac{E-\epsilon}{2E^2} [1+2N_E] \;.
\label{lambda}
\end{equation}
We have added and subtracted $[1+2N_E]\ [\tm \epsilon /2E^2] $ to (\ref{spdm_2})
to ensure that (i) both $\tilde{\rho}({\bf r}) $ and $\Lambda ({\bf r})$ are free from ultra-violet and infra-red divergencies in all dimensions, and (ii) that only small momenta $k<k_0$ contribute to the phase correlator $\Lambda ({\bf r})$.

By comparing (\ref{lambda}) and  (\ref{Xi_low}) we see that they coincide at large distances
up to leading terms. It means that in 2D and in 1D at zero temperature, equation (\ref{spdm_3})
can be trusted up to exponentially large scales, while in 1D at finite temperature
$T \ge \vert \tm \vert$ it works at least up to distances $\sim n_{\rm s}/mT$.

We are now in position to define the quasi-condensate density as the limiting value
of $\tilde{\rho}({\bf r}\to \infty )$:
\begin{equation}
n_{\rm qc} = n- \sum_{\bf k} \,
\frac{\epsilon }{E^2}
\Big[ \frac{E-\epsilon +\tm }{2} +(E-\tm )N_E \Big] \, .
\label{nquasi}
\end{equation}
The physical meaning of this relation is the amplitude squared of the order parameter field
at large distances. There is a certain degree of freedom in attributing terms which do not result
in the power-law or exponential decay of $\rho ({\bf r})$ to $d({\bf r})$ or to $\Lambda ({\bf r})$.
The idea behind our choice is three-fold: (i) equations(\ref{spdm_3}), (\ref{ntilde}), and (\ref{lambda}) are
valid as written in all spatial dimensions; (ii) at large distances $\Lambda ({\bf r})$
has the structure of hydrodynamic phase correlations; (iii) the final expressions have the
simplest form possible within the same accuracy.
It is also worth mentioning that $n_{\rm qc}$ is a quantity which controls all physical processes
happening in the WIGB at short distances at low temperature, e.g.\ $m$-body recombination rates.
In all spatial dimensions it plays the same role as the condensate density in 3D system as long as
one is interested in length scales not much larger than the healing length \cite{Kagan87,Svistunov}.

Finally, the condensate density is defined from
\begin{equation}
n_{0}\,  =\, n_{\rm qc} \, {\rm e}^{-\Lambda (\infty )}\, .
\label{ncnd}
\end{equation}
It is not accidental that the ultimate long-wave length property of the superfluid system
is determined last. It was always an unpleasant feature of numerous mean-field treatments that
the crucial parameter determining physics at short scales was linked to $n_0$.

\section{Normal region}
\label{sec:Normal_region}

As was mentioned in the Introduction,  we assume that the temperature is
not very high, so that  the condition (\ref{lambda_T_vs_R_0}) is preserved and  the quantity $U$
remains the only parameter characterizing the interaction between the particles.

\subsection{Thermodynamic functions}

In the normal region, where there are no anomalous correlators, we have to deal only with
the Dyson equation for the Green's function $G$ in terms of self-energy $\Sigma_{11}$. Within the leading-order approximation, $\Sigma_{11}=2nU$ (in 2D this formula implies an adequate choice of the parameter $\epsilon_0\equiv \epsilon_0(T)$ for the effective interaction $U$, discussed in subsection \ref{subsec:2D_large_T}),  and the expression for $G$ yields the self-consistent
relation for the number density,
\begin{equation}
n\, =\,  \sum_{\bf k} \, \left[{\rm
e}^{\tilde{\epsilon}(k)/T} - 1 \right]^{-1} \, , \label{n_normal}
\end{equation}
with
\begin{equation}
\tilde{\epsilon}(k)\, =\, \epsilon(k) - \tm \, \equiv \, \epsilon(k) + |\tm|  \; . \label{eff_e}
\end{equation}
For the pressure we find
\begin{eqnarray}
p\, =\, \int_{-\infty}^{\tm} n\,  { {\rm d} \mu \over {\rm d} \tm}
\, {\rm d} \tm  \, =\, \int_{-\infty}^{\tm} n\, \left(
1+2U{{\rm d} n \over {\rm d} \tm} \right) \,{\rm d} \tm\nonumber \\
\, =
\int_{-\infty}^{\tm} n\, {\rm d} \tm \, +\,  n^2U \; .
\label{p_norm}
\end{eqnarray}
Using
\begin{equation}
n\, = \, \sum_{\bf k} \, N_{\tilde{\epsilon}}
\;\equiv  -T {{\rm d}  \over {\rm d} \tm} \, \sum_{\bf k} \, \ln \left[1-{\rm e}^{-\tilde{\epsilon}(k)/T} \right]
\; , \label{obs}
\end{equation}
we finally arrive at
\begin{equation}
p\, = \,n^2 U -T \sum_{\bf k} \, \ln
\left[1-{\rm e}^{-\tilde{\epsilon}(k)/T} \right] \; .
\label{p_normal}
\end{equation}
Equations (\ref{n_normal}), (\ref{eff_mu}), and (\ref{p_normal})
specify $n$, $\mu$, and $p$ as functions of $(T,\tm)$.
Utilizing  (\ref{s2}), with $x\equiv \tm $, and then (\ref{e1}), we find for the entropy
\begin{equation}
s=\sum_{\bf k} \, \left\{{\tilde{\epsilon}(k)\over T}N_{\tilde{\epsilon}}
-\ln\left[1-{\rm e}^{-\tilde{\epsilon}(k)/T}\right] \right\} , \label{s_nrml}
\end{equation}
and energy
\begin{equation}
\varepsilon=Un^2+\sum_{\bf k} \, \epsilon (k)N_{\tilde{\epsilon}} \, .
\label{e_nrml}
\end{equation}

\subsection{Effective interaction in the normal regime in 2D}
\label{subsec:2D_large_T}

\begin{figure}
\includegraphics[angle=-90, scale=0.2, bb=-30 -500  600 100]{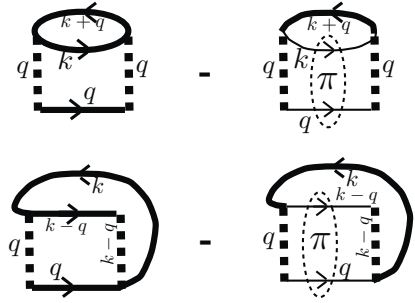}
\caption{ Second-order contributions to the self-energy in the normal regime, including the sunrise diagram.}
\label{fig: sunrise}
\end{figure}

In the quasi-condensate region, the parameter $\epsilon_0$ defining the effective interaction $U$
can vary over a wide range of values thanks to the retained second-order correction that produces counter-terms
compensating for the arbitrary choice. Since in the normal region we confine ourselves to the first-order expression for the self-energy, $\Sigma=2nU$,
we need to choose the value of $\epsilon_0$ which minimizes the omitted second-order contributions
originating from the sunrise diagram, see figure~\ref{fig: sunrise}.
Due to momentum independence of the pseudo-potential line, the third (fourth) diagram is identical to the first (second) one; we thus consider only the first two diagrams and multiply the result by a factor of 2. The parameter $\epsilon_0$ can be much smaller or much larger than $T$. In either case the leading term comes from the logarithmic ultraviolet contribution from the product of two propagators, yielding the result

\begin{equation}
\Sigma^{(2)} \, =\, -2nU^2 \sum_{k>k_T} \, \left[ \Pi (k) - {m\over k^2}\right] + nU{\cal O}(mU)\, ,
\label{sec_order_sig}
\end{equation}
where $k_T$ is the thermal momentum. This leads to the expression
\begin{equation}
\Sigma \, =\, 2nU\left[ 1+ {mU\over 4\pi} \ln {T\over \epsilon_0}  + {\cal O}(mU) \right] \, .
\label{sec_order_sig2}
\end{equation}
Comparing the first two terms in the brackets with (\ref{C_12_2D}) and  (\ref{U1_U2_expd}), we see that if we take the value of $\epsilon_0$ substantially away from $T$, then the leading
second-order correction to the expression $\Sigma = 2nU$ amounts to renormalizing the value of the pseudo-potential $U$ in such a way that it corresponds to $\epsilon_0\sim T$. This brings us to the conclusion that $\epsilon_0 \sim T$ is the optimal choice for $\epsilon_0$, in which case omitting the diagrams shown in figure~\ref{fig: sunrise} is justified by the parameter  $mU \ll 1$.

\subsection{Expansion parameter}
\label{subsec:exp_par_norm}

At  temperatures above the fluctuation region where the renormalized chemical potential
remains small $\vert \tm \vert \ll T$ the  expansion parameter is defined by the infra-red behaviour of the zero-frequency lines. The  dimensionless factor, $\gamma_T$, associated with adding an extra full interaction vertex to a diagram is given by (\ref{full_T_rel1}), but now with $G \sim T/\tilde{\epsilon}(k)$, resulting in the estimate
\begin{equation}
\gamma_T\, \sim \, \frac{TUm^{d/2}}{\tm^{2-d/2}} \ \sim \
\left( \frac{ \tm^{\rm fluc} }{ \tm } \right) ^{2-d/2} \, ,
\label{normal_rel2}
\end{equation}
where $\tm^{\rm fluc}$ is the size of the fluctuation region in terms of the chemical potential.
At temperatures much higher than the  degeneracy temperature $T_{\rm c} \sim n^{2/d}/m$ where
$\tm \approx -T \ln (n\lambda_T^d) $, 
all corrections are small in the parameter
\begin{equation}
\gamma_T\, \sim \,  \frac{U}{\vert \tm \vert \ \lambda_T^d  } \, .
\label{normal_rel2b}
\end{equation}

\section{Pseudo-Hamiltonian}
\label{sec:Pseudo}

In Section~\ref{sec:Quasicondensate} we have obtained the thermodynamics of the system
in the quasi-condensate region, see (\ref{n_total_qc})-(\ref{n_prime}), (\ref{p3}), (\ref{s_cnd}),  and (\ref{e_cnd}) which specify $n$, $\mu$, $p$, $s$, and $\varepsilon$ as functions of two independent variables, $\tm$ and $T$, in the form of integrals involving the $\tm$-dependent quasiparticle spectrum, $E(k)$, given by (\ref{E2}). So far, we did not discuss the physical meaning of the function $E(k)$, since this was not necessary for our derivations. On the other hand, equation (\ref{s_cnd}) for the entropy clearly suggests that, within our approximation, the system is equivalent to a system of non-interacting bosonic quasiparticles with dispersion $E(k)$, described by the Hamiltonian
\begin{equation}
\tilde{H}=E_0+\sum_{\bf k}\, E(k)\, \hat{n}_{\bf k}\, ,
\label{H_ps}
\end{equation}
where $\hat{n}_{\bf k}$ is the quasiparticle occupation number operator and $E_0$ is some constant. The natural question is then whether the Hamiltonian (\ref{H_ps}) can be used for calculating
other thermodynamic quantities within the non-interacting quasiparticle gas picture.
By direct comparison with (\ref{n_total_qc})-(\ref{n_prime}),  (\ref{p3}), (\ref{s_cnd}),  and (\ref{e_cnd}), it can be shown that this is indeed the case and one can write
\begin{equation}
E-\mu N=\langle\tilde{H}\rangle=E_0+\sum_{\bf k}E(k)N_E\, ,
\label{H_ps1}
\end{equation}
\begin{equation}
\Omega = -pV=-T\ln\left( {\rm  Tr\, e}^{-\tilde{H}/T}\right)  .
\label{omega_ps}
\end{equation}
It is important, however, to remember that $E_0$ is a function of both $\tm$ and $T$,
satisfying
\begin{eqnarray}
E_0=  - \frac{\tm^2}{2U}+2\tm n'-Un'^2 + I_1^{(d)}(\tm) \, .
\label{E_0_qc}
\end{eqnarray}
In view of this temperature dependence, we refer to (\ref{H_ps}) as a  {\it pseudo}-Hamiltonian,
rather than an effective Hamiltonian.

The pseudo-Hamiltonian description can be applied to the normal region as well.  In this case,
\begin{equation}
E_0=-n^2U \, ,
\label{E_0_normal}
\end{equation}
were $n\equiv n(\tm,T)$ is specified by (\ref{n_normal}). The  $\tm$-dependent
spectrum of quasi-particles is now given by (\ref{eff_e}).

\section{Fluctuation region}
\label{sec:fluct}

In the fluctuation region the
long-wave part of the classical-field component of the quantum field
experiences strong (non-perturbative) fluctuations. It is exclusively due to
these fluctuations that the diagrammatic technique for the quantum field
loses an expansion parameter.  The special role of
the classical-field component is evident from the fact that on  approach
to the fluctuation region the leading contribution to the expansion parameter $\gamma_T$
is associated with zero-frequency propagators, see subsections \ref{subsec:finite_T} and \ref{subsec:exp_par_norm}. The diagrammatic technique built on zero-frequency propagators (with perturbative parts of self-energies included) and the interaction represented by pseudo-potential $U$ corresponds to the
diagrammatic expansion of the Gibbs distribution
\begin{equation}
Z=\int {\rm e}^{-H[\psi]/T} {\cal D}\psi
\label{Z_classical}
\end{equation}
for the in momentum space truncated classical field
\begin{equation}
\psi ({\bf r}) = \sum_{k<k'} \psi_{\bf k}\, {\rm e}^{ {\rm i} {\bf k}\cdot{\bf r}}\, ,
\label{truncate}
\end{equation}
with the Hamiltonian functional
\begin{equation}
H[\psi] = \int \left[ {1\over 2m} |\nabla \psi|^2 +{U\over 2}|\psi|^4 -\mu' |\psi|^2 \right]\, {\rm d}{\bf r}\, .
\label{H_classical}
\end{equation}
Here $k'\ll k_T=\sqrt{mT}$ is a truncation momentum (to be discussed later), and $\mu'\equiv \mu'(k')$ is the reduced chemical potential obtained from the bare one by subtracting relevant perturbative contributions  from all the harmonics with $k>k'$, including the quantum ones. The bottom line is that describing the quantum gas in the fluctuation region reduces to solving the classical-field problem (\ref{Z_classical})-(\ref{H_classical}), in its fluctuation region.

In $d=2,3$,  numerical solutions to the problem (\ref{Z_classical})-(\ref{H_classical})---in terms of scaling functions for  thermodynamic quantities---are available in the literature  and have already been used to accurately describe weakly interacting 2D and 3D quantum gases  in the fluctuation region \cite{Tc3D,Arnold,Tc2D}. The same is possible in 1D, but to the best of our knowledge it has not been done so far.

In the fluctuation region, an important quantity is the momentum $\tilde{k} \ll k_T$ separating weakly and
strongly coupled classical modes. For a quantitatively accurate description of the fluctuating classical-field sub-system,  the momentum $k'$ separating classical modes of interest from the rest of the modes has to be much larger than $\tilde{k}$. However, as long as we are interested in the order-of-magnitude estimates, it is safe and convenient to set $k'\sim \tilde{k}$, so that all the modes we are dealing with in
(\ref{Z_classical})-(\ref{H_classical}) are strongly coupled. To estimate $\tilde{k}$ we note that
with $k'\sim \tilde{k}$ all three terms in the Hamiltonian (\ref{H_classical}) have to be of the
same order of magnitude, that is
\begin{equation}
\tilde{k}^2/m \, \sim \, | \tm | \, \sim \, \tilde{n} U \; ,
\label{est1}
\end{equation}
where
\begin{equation}
\tilde{n} \sim  \sum_{k<\tilde{k}} |\psi _k |^2 = \sum_{k<\tilde{k}} n_k \; ,
\label{tilde_n}
\end{equation}
is the long-wavelength contribution to the total density, and
$\tm \equiv \mu'(k'\to \tilde{k})$. For $\tilde{n}$ we have $\tilde{n} \sim \tilde{k}^d n_{\tilde{k}}$, and since
$\tilde{k}$ is separating strongly coupled long-wave harmonics from
slightly perturbed  short-wave ones, the order-of-magnitude
estimate for $n_{\tilde{k}}$ follows by continuity from the ideal system
formula:
\begin{equation}
n_{\tilde{k}} \sim \frac{T}{\tilde{k}^2/2m-\tm}
\sim \frac{T}{ | \tm | }  \; .
\label{reltnc}
\end{equation}
Substituting this back into (\ref{est1})-(\ref{tilde_n}) yields
\begin{equation}
\tilde{k} = (m^2 T U)^{1\over 4-d} \, ,
\label{est2_k}
\end{equation}
\begin{equation}
\tilde{n} \sim (m^d T^2 U^{d-2})^{1\over 4-d}   \, ,
\label{est2_n}
\end{equation}
\begin{equation}
|\tm | \sim (m^d T^2 U^2)^{1\over 4-d}  \, .
\label{est2_mu}
\end{equation}
The estimates (\ref{est2_n})-(\ref{est2_mu}) imply the following parameterization of the grand-canonical equation of state in the fluctuation region and its vicinity:
\begin{equation}
n=n_{\rm c}^{(d)}(T)+ (m^d T^2 U^{d-2})^{1\over 4-d} \lambda^{(d)}(x) \, ,
\label{param_n}
\end{equation}
\begin{equation}
x={\mu -\mu_{\rm c}^{(d)}(T) \over (m^d T^2 U^2)^{1\over 4-d} } \, .
\label{param_mu}
\end{equation}
Here $\lambda^{(d)}(x)$ is a dimensionless scaling function of a dimensionless scaling variable $x$. In 2D and 3D, the quantities $n_{\rm c}^{(d)}(T)$ and $\mu_{\rm c}^{(d)}(T)$ are the critical values of density and  chemical potential at a given temperature.
In 1D, where the phase transition is absent, one can set, without loss of generality, $\mu_{\rm c}\equiv 0$, and, correspondingly, $n_{\rm c}(T)\equiv n(\mu \! =\! 0,T)$.

Similarly, superfluid and condensate densities---in the dimensions in which they are meaningful---are parameterized as
\begin{equation}
n_{\rm s}= (m^d T^2 U^{d-2})^{1\over 4-d} f_s^{(d)}(x) \qquad (d=2,3) \, ,
\label{param_n_s}
\end{equation}
and
\begin{equation}
n_0= m^3 T^2 U f_0(x) \qquad (d=3) \, .
\label{param_n_0}
\end{equation}

The functions
$\lambda(x)$, $f_s(x)$, and $f_0(x)$ are universal for all weakly interacting U(1)-symmetric systems in the given dimension (no matter quantum or classical, continuous space or lattice); the numerical data for them is available for $d=2,3$ in  \cite{Tc3D,Tc2D}. By numerically solving the problem (\ref{Z_classical})-(\ref{H_classical}), it was established that, 
up to higher-order corrections in the parameter $\gamma_0$, \cite{Tc3D,Arnold,Tc2D}
\begin{equation}
n_{\rm c}^{(3D)} = n_{\rm c}^{(0)}(T) - C  m^3T^2U \, ,
\qquad C=0.0142(4) \, ,
\label{n_c_3D}
\end{equation}
\begin{equation}
n_{\rm c}^{(2D)} = { mT\over 2\pi} \ln \left( {\xi\over mU} \right) \, , \qquad \xi=380\pm 3 \, ,
\label{n_c_2D}
\end{equation}
with $n_{\rm c}^{(0)}(T)$ the critical density for Bose-Einstein condensation in an ideal 3D gas.
Analogous relations for $\mu_{\rm c}^{(d)}(T)$ read \cite{Tc3D,Arnold,Tc2D}
\begin{equation}
\mu_{\rm c}^{(3D)} = 2U n_{\rm c}^{(0)}(T) +{m^3T^2U^2\over \pi^2}\ln \left( {0.4213(6)\over \sqrt{m^3TU^2}}\right)\, .
\label{mu_c_3D}
\end{equation}
\begin{equation}
\mu_{\rm c}^{(2D)} = { mT U\over \pi} \ln \left( {\xi_{\mu} \over mU} \right) \, , \qquad \xi_{\mu}=13.2\pm 0.4 \,  ,
\label{mu_c_2D}
\end{equation}

Note that the leading terms in the above relations, as well as order-of-magnitude estimates for  the values of sub-leading terms, are readily obtained from the condition $\gamma_T\sim 1$; the accurate numerical treatment of the problem (\ref{Z_classical})-(\ref{H_classical}) is necessary to fix the values of the dimensionless constants.

By re-writing (\ref{n_c_2D}), (\ref{n_c_3D}) in terms of the critical temperature
as a function of density, we get
\begin{equation}
T_{\rm c}^{(2D)} = {2\pi n\over m\ln (\xi/mU)}\, ,
\label{Tc_2D}
\end{equation}
\begin{equation}
T_{\rm c}^{(3D)} = T_{\rm c}^{(0)}(1+C_0an^{1/3})\, , \qquad C_0=1.29\pm 0.05 \, ,
\label{Tc_3D}
\end{equation}
where $T_{\rm c}^{(0)}$ is the critical temperature of the ideal 3D gas. Now it is easy to estimate $\tilde{k}$. The relevant temperature for having strong fluctuations is given by
\begin{equation}
T_{\rm fluct}\approx T_{\rm c} \quad (d=2,3) \, , \qquad T_{\rm fluct} \sim nU/\gamma_0 \quad (d=1) \, .
\label{rlv_T}
\end{equation}
The estimate (\ref{est2_k}) at $T=T_{\rm fluct}$ then yields
\begin{equation}
\tilde{k} / k_{T_{\rm fluct}} \, \sim \, \left\{ {\begin{array}{*{20}c}
{ \gamma_0^{2/3} \qquad (d = 3)\, , } \\
  { \gamma_0^{1/2} \qquad (d = 2)\, , }  \\
    { \gamma_0^{1/2} \qquad (d = 1)\, . }  \\
\end{array}} \right.
\label{k_tilde_over_k_T}
\end{equation}
We see that $\tilde{k} \ll k_{T_{\rm fluct}}$ in all cases, which justifies the statement that the physics of the fluctuation region is that of a (strongly interacting) classical field.

Putting aside $n_{\rm s}$ and $n_0$, which are most sensitive to non-perturbative fluctuations of the classical field $\psi$, the next two quantities which are sensitive to fluctuations, especially in 1D and 2D, are the density and the chemical potential---see estimates (\ref{n_tilde_mu_tilde}).
For other quantities---as is seen from (\ref{p_e_s_fluct})---neglecting  fluctuation corrections will not result in a significant error. It is also important that up to a few
dimensionless constants characterizing sub-leading contributions to critical values of $p$, $\varepsilon$, and $s$, the fluctuation corrections to these quantities are expressed in terms of the same function $\lambda^{(d)}(x)$ and its integral. Indeed, equation (\ref{rel1}) implies the following relation for $p$:
\begin{eqnarray}
p=p_{\rm c}^{(d)}(T)+ n_{\rm c}^{(d)}(T)[\mu-\mu_{\rm c}^{(d)}(T)]\nonumber \\
+\,  (m^{2d} T^4 U^d)^{1\over 4-d}\int_0^x  \lambda^{(d)}(x')\, {\rm d} x' \, ,
\label{param_p}
\end{eqnarray}
and then with (\ref{s2})-(\ref{e1}) one obtains similar relations for $s$ and $\varepsilon$.

\section{Comparison with numerical results}
\label{sec:Numerics}

\begin{figure}
\includegraphics[angle=-90, scale=0.3, bb=-50 -270 560 330]{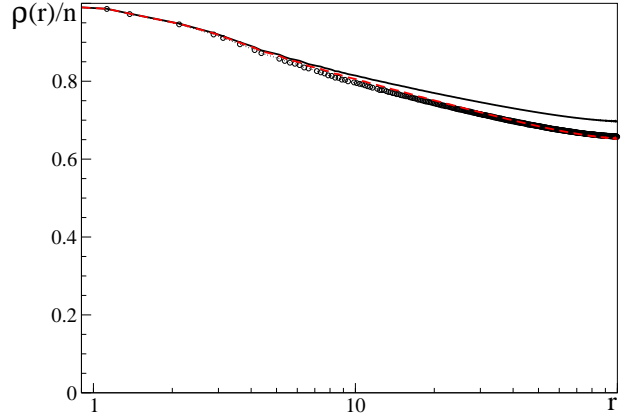}
\caption{ (Colour online) Density matrix for the two-dimensional Bose-Hubbard model (\ref{BH}) at
$U/t=0.25$, $T/t=0.5$, and $n=0.5$ with periodic boundary conditions on a square lattice lattice
with $200 \times 200$ sites. Theoretical lines were obtained from (\ref{spdm_2}) (dashed red line)
and (\ref{spdm_3}) (solid black line). }
\label{fig:rho2D}
\end{figure}

\begin{figure}
\includegraphics[angle=-90, scale=0.3, bb=-50 -270 560 330]{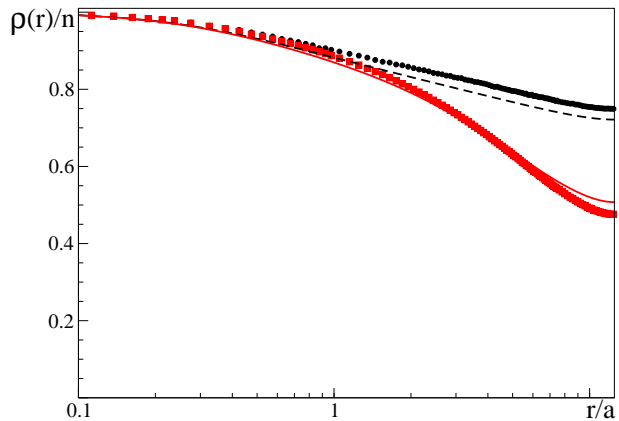}
\caption{ (Colour online) Density matrix for the one-dimensional Bose gas with $na=8$ at temperature
$mT/(2\pi n^2)=0.001$ (black circles) and $mT/(2\pi n^2)=0.004$ (red squares) on a circle of length
$L/a=25$. Theoretical curves are represented by (\ref{spdm_3}), (\ref{ntilde}), and  (\ref{lambda}) with the dashed black line for the lower temperature and the solid red line for higher one. }
\label{fig:rho1D}
\end{figure}

To see how accurate our description is for the density matrix
we performed Monte carlo simulations of weakly interacting
two- and one-dimensional systems where phase fluctuations are the strongest.
We deliberately aim at systems with substantial decay of the density matrix to
see the difference between the perturbative treatment and exact solution.
In figure~\ref{fig:rho2D} we present data for the two dimensional square-lattice system
described by the Bose-Hubbard Hamiltonian
\begin{equation}
H = -t \sum_{<{\bf r}{\bf r}'>} \left( \psi^{\dag}_{{\bf r}' } \psi_{ {\bf r} } +h.c. \right)
+\frac{U}{2}  \sum_{{\bf r}} n_{\bf r} [n_{\bf r}  -1]  \;,
\label{BH}
\end{equation}
where $t$ is the hopping amplitude between the nearest neighbor sites. Model (\ref{BH}) was simulated
for $U/t=0.25$, $T/t=0.5$ and filling factor $n=0.5$ on a lattice with $L \times L = 200 \times 200$ points
by using the Worm Algorithm approach \cite{latticeworm}. Formally, all expressions in this manuscript
relating energy spectrum, particle density, and density matrix remain valid for an arbitrary dispersion relation with $\epsilon (k) \to k^2/2m $ at small momenta. Moreover, the theory is supposed to work as written for systems with periodic boundary conditions provided the sums over momenta are understood as $ L^{-d} \sum_{{\bf k} \ne 0} $, where
${\bf k} = 2\pi {\bf n}/L$ and  ${\bf n}$ is a $d$-dimensional integer. To suppress statistical noise,
simulation data for $\rho({\bf r})$ were collected to spherically symmetric bins. We are using
exactly the same procedure to present theoretical data. Thus, the only difference between
the essentially exact (up to statistical errors) simulation data and the theory is due to the finite
value of $U$.
Even for relatively large values of $U$ and temperature $T \sim 0.6T_{\rm c}$ we observe
a remarkable accuracy of (\ref{spdm_2}). One can get an idea of the systematic theoretical
error by comparing curves derived from (\ref{spdm_2}) and (\ref{spdm_3}). [The latter is obtained by exponentiating the phase correlator at large distances].

In figure~\ref{fig:rho1D} we apply our theory to the Lieb-Liniger model of the one-dimensional
Bose gas with contact interactions. Since exact correlation functions at finite temperature are
not known, we performed Monte carlo simulations using recently developed techniques for
continuous space systems \cite{PREworm,trentoworm}. It is convenient to introduce the on-dimensional characteristic length $l_0=2/mU$ as the unit of length. The limit of weak interactions
is obtained then by considering large densities $na \gg 1$, and we have chosen $nl_0=8$ corresponding
to the power law-decay of off-diagonal correlations at zero temperature with the exponent
$\alpha = 1/\pi \sqrt{2nl_0}=1/4\pi $. The upper curve in figure~\ref{fig:rho1D}
shows results for low temperature $ mT/(2\pi n^2) = 0.001$, or $T/nU\approx 0.025$ when
thermal phase fluctuations are barely influencing the data for the simulated system size
$L/l_0=25$. When temperature is increased to $ mT/(2\pi n^2) = 0.004$ 
(lower curve in figure~\ref{fig:rho1D}) we clearly see the bimodal
decay of the density matrix and the crossover from the power-law to exponential decay.
Contrary to the 2D case, equation (\ref{spdm_3}) captures the actual behaviour better than
(\ref{spdm_2}) which is not so surprising given the large suppression of $\rho ({\bf r})$ at large
distances.

\begin{figure}[tbp]
\includegraphics[scale=0.6, bb=-70 80 500 450]{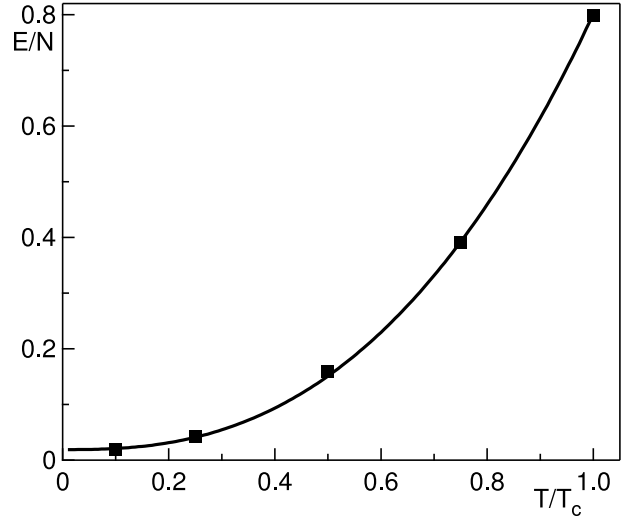}
\caption{ Energy per particle in units of $T_{\rm{c}}$ for the three-dimensional Bose gas with $na^3=10^{-6}$ vs. temperature, in the quasi-condensate region. Squares are quantum Monte Carlo results, solid line is the theoretical curve  (\ref{e_cnd}).}
\label{fig:energyqc_comparison}
\end{figure}

\begin{figure}[tbp]
\includegraphics[scale=0.6, bb=-70 80 500 450]{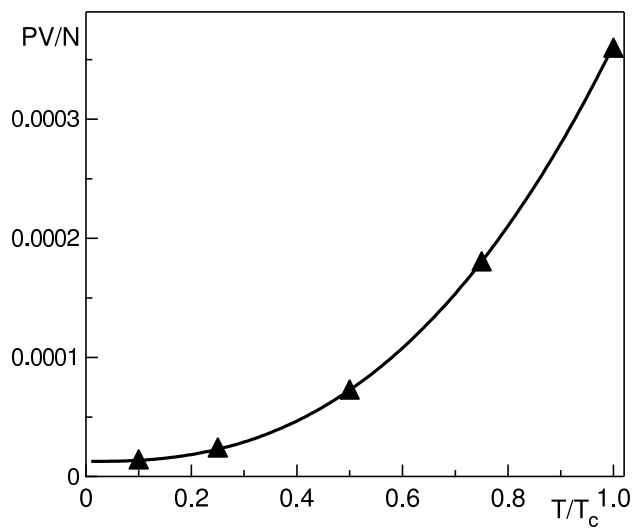}
\caption{ Pressure in units of $1/2ma^2$ for the three-dimensional Bose gas with $na^3=10^{-6}$ vs. temperature in the quasi-condensate region. Triangles are quantum Monte Carlo results, solid line is the theoretical curve  (\ref{p3}).}
\label{fig:pressureqc_comparison}
\end{figure}

\begin{figure}[tbp]
\includegraphics[scale=0.55, bb=-100 80 500 450]{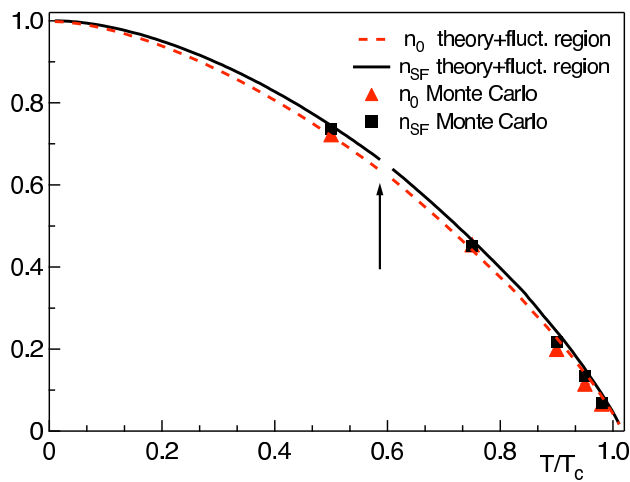}
\caption{ Quantum Monte Carlo results for superfluid fraction (squares) and condensate fraction (triangles) for the three-dimensional Bose gas with $na^3=10^{-6}$. Solid and dashed lines are a combination of theoretical expressions from (\ref{nslandau}) for the superfluid density and (\ref{n_total}) for condensate density with preexisting classical Monte Carlo results~\cite{Tc3D} for the fluctuation region (the arrow is pointing to the gap in the curves where the two results are matched).}
\label{fig:nSF_ncnd_comparison}
\end{figure}

\begin{figure}[tbp]
\includegraphics[scale=0.55, bb=-100 80 500 480]{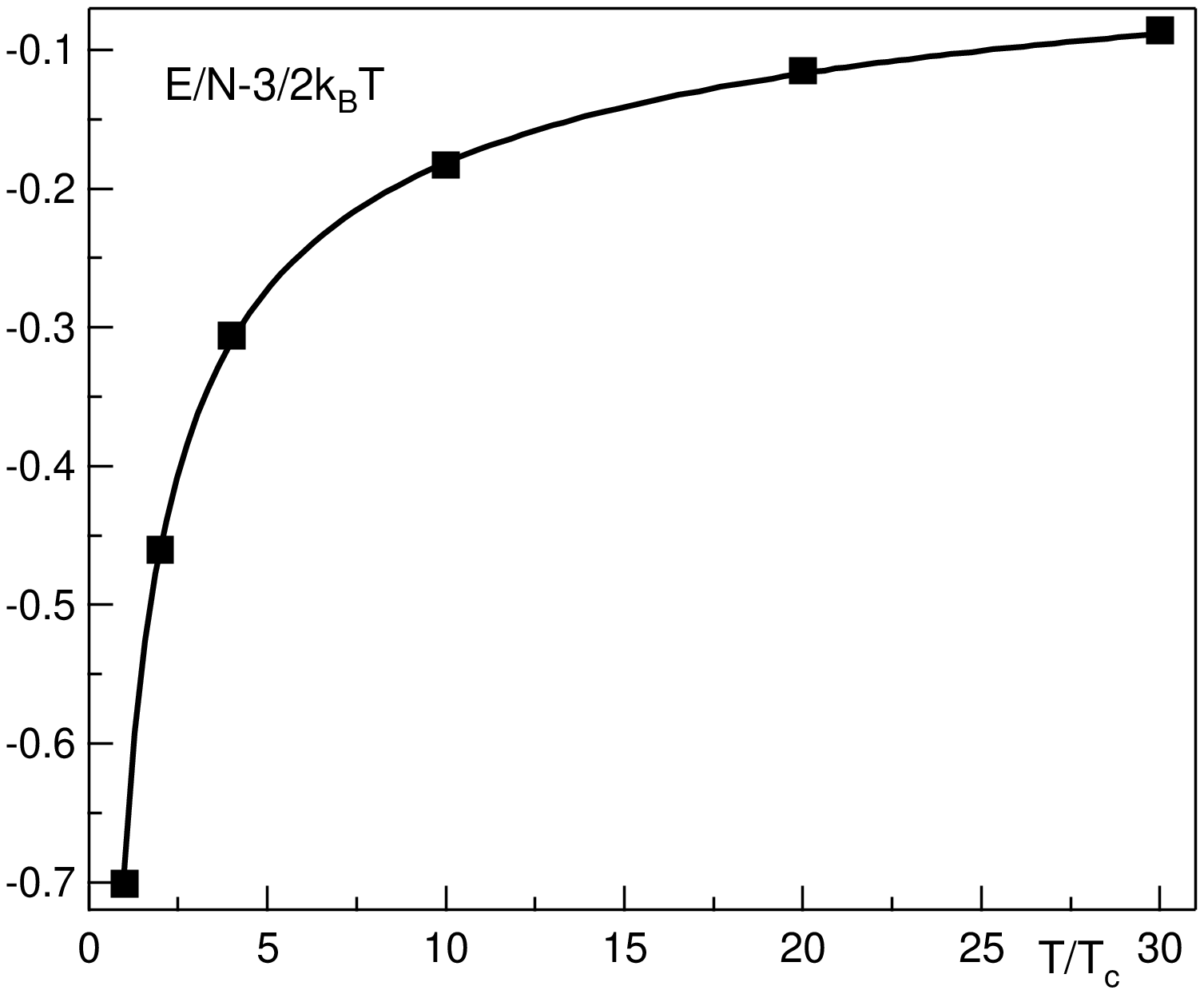}
\caption{Deviation of the energy per particle (in units of $T_{\rm{c}}$) from the classical law for the three-dimensional Bose gas with $na^3=10^{-6}$ vs. temperature, in the normal region. Squares are quantum Monte Carlo results, solid line is the theoretical curve  (\ref{e_nrml}).}
\label{fig:energyn_comparison}
\end{figure}

\begin{figure}[tbp]
\includegraphics[scale=0.6, bb=-50 80 500 450]{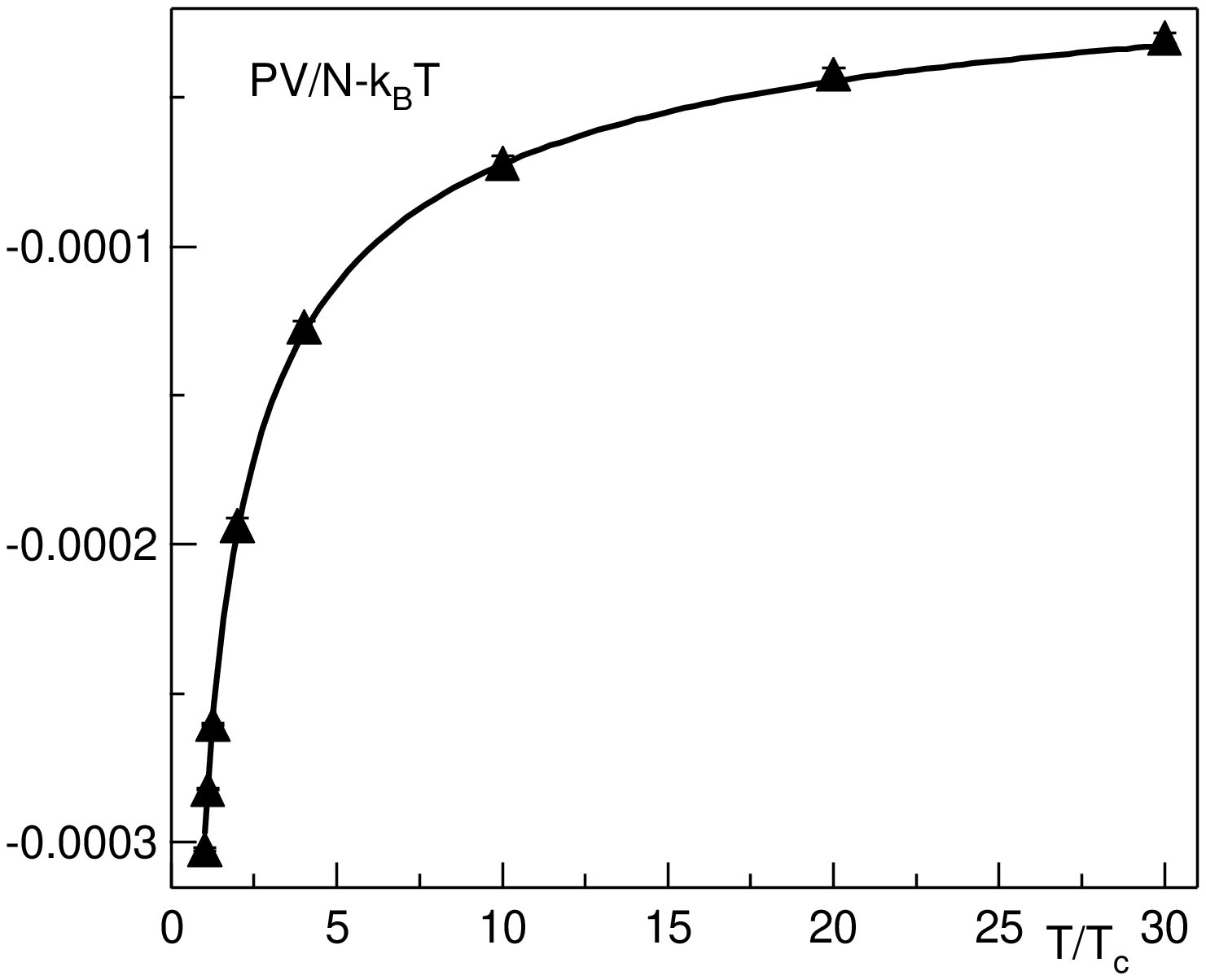}
\caption{Deviation of the pressure (in units of $1/2ma^2$) from the classical law for the three-dimensional Bose gas with $na^3=10^{-6}$ vs. temperature, in the normal region. Triangles are quantum Monte Carlo results, solid line is the theoretical curve  (\ref{p_normal}).}
\label{fig:pressuren_comparison}
\end{figure}

Finally,  figures \ref{fig:energyn_comparison} and \ref{fig:pressuren_comparison}  show the agreement of Monte Carlo results for the energy (squares) and the pressure (triangles) at temperatures $T>T_{\rm c}$ with theoretical curves from (\ref{e_nrml}) and (\ref{p_normal}) (solid lines).

In the following we compare thermodynamic functions obtained in Sections \ref{sec:Quasicondensate} and \ref{sec:Normal_region} with exact quantum Monte Carlo results of a weakly interacting three-dimensional system. Let us start with the case of a homogeneous system with a small parameter $na^3= 10^{-6}$. In figures \ref{fig:energyqc_comparison} and \ref{fig:pressureqc_comparison} we compare Monte Carlo results for the energy (squares) per particle and pressure (triangles) at temperatures $T<T_{\rm c}$, with the theoretical curves from (\ref{e_cnd}) and (\ref{p3}) (solid lines). The analytical expressions are in very good agreement with numerical results up to temperatures $T\sim T_{\rm c}$ and there is no need to switch to the description in terms of universal functions \cite{Tc3D}. This is consistent with the estimate (\ref{p_e_s_fluct}) showing that fluctuation contributions to energy and pressure are very small.
The same does not apply to the superfluid and condensate densities for which the perturabtive expansion in $U$ is not valid as fluctuations of the order parameter cannot be neglected on approach to the critical temperature. In figure~(\ref{fig:nSF_ncnd_comparison}) squares and triangles are the results of quantum Monte Carlo simulations for the superfluid and condensate fraction, respectively. Solid and dashed lines are the theoretical expressions from (\ref{nslandau}) for the superfluid density and (\ref{n_total}) for the condensate density, respectively,
combined with preexisting classical Monte Carlo results~\cite{Tc3D} for the fluctuation region. At a temperature $T\gtrsim 0.6T_{\rm c}$ we switch to the universal description, as $\gamma^{\rm (full)}_T$ [see (\ref{gamma_T_rel3})] becomes of the order of unity.
For $T\sim T_{\rm c}$ the remaining discrepancy between classical and quantum Monte Carlo results is due to finite interaction strength. One should not be misled by the small value of the gas parameter in the fluctuation region
since the proper parameter controlling the size of the fluctuation region in temperature, $\Delta T/T_{\rm c}$, 
and the accuracy of the universal description involves a large numerical prefactor and is rather 
$na^3/10^{-5}$, see \cite{Tc3D}.

\section{Conclusions}

We have shown that Beliaev's diagrammatic technique regularized by adding a small term  explicitly breaking the U(1) symmetry to the Hamiltonian yields a simple and controllable description of the weakly interacting Bose gas in any dimension. This approach is especially convenient for obtaining thermodynamic functions that are not sensitive (up to higher-order corrections) to the long-range phase fluctuations.
The symmetry breaking term introduces a small gap in the otherwise gapless 
Goldstone mode and suppresses long-range fluctuations of the phase of the order
parameter, thereby introducing a genuine condensate density
and thus substantially simplifying the description of lower-dimensional
systems. In this sense, the symmetry-breaking trick is an alternative to
Berezinskii's finite-size trick and Popov's special (hydrodynamic) treatment of
long-wave parts of the fields. At the end of the calculation, the symmetry breaking term
is set to zero.

Another important feature of our approach is that it completely avoids such notions as
seed- or quasi-condensate for its construction. The quasi-condensate density defined as the
modulus squared of the order parameter field at distances larger then the healing length is
calculated as the long-wave property of the theory, rather than serving as an input parameter.
It is only natural to have the theory which handles the short-range physics first and
uses it for dealing with the low-energy physics next. As far as  long-range off-diagonal 
correlations are concerned, the approach  produces
accurate answers for the off-diagonal correlation functions up to distances where further evolution
of the correlators is controlled by generic hydrodynamic relations, and thus can be accurately extrapolated to arbitrarily large distances.

We confined our analysis to the most typical and non-trivial (in 2D and 3D) case of dilute gas, when the size of the potential $R_0$ is much smaller than 
the  distance between the particles. In the (Boltzmann) high-temperature regime, we have restricted ourselves to low enough temperatures at which  de Broglie 
wavelength  remains much larger than $R_0$ and the answers do not depend on the details of the pair potential. Under the  two above-mentioned 
conditions, the effective inter-particle interaction is described by a single parameter $U$. Technically, generalization of the theory to the case of $R_0 \gtrsim n^{-1/d}$ is 
straightforward. In this case, the weakness of interaction literally means Born interaction potential, and complete summation of ladder diagrams in 2D 
and 3D becomes irrelevant.  The description of a normal gas at $\lambda_T \lesssim n^{-1/d}$ is a well-studied  topic that goes beyond the scope of the present paper.

The pair of variables $\tm=\mu-2nU$ and $T$ form the most convenient set of parameters determining the rest
of thermodynamic quantities, including the condensate density, if any. 
The structure of the answers corresponds to a picture of non-interacting bosonic quasiparticles with the Bogoliubov spectrum (\ref{E2}), on top of the `vacuum' with temperature-dependent energy. This allows one to introduce the bilinear bosonic pseudo-Hamiltonian (\ref{H_ps}) with Bogoliubov spectrum. We argue that within the bilinear pseudo-Hamiltonian ansatz our results cannot be further improved at least at $T\gtrsim nU$. In the $T \ll nU$ limit, when thermodynamics is exhausted by dilute non-interacting phonons, using the improved value of the sound velocity obtained directly from the zero-temperature compressibility, yields more accurate answers for thermodynamic quantities, but is inconsistent with retaining the Bogoliubov form of the spectrum for {\it all} the momenta. We also show that any attempt to improve the self-consistent mean-field description of the superfluid region, aimed at getting rid of the spurious first-order phase transition, is senseless as long as the fluctuation-induced non-perturbative shift of the critical temperature is not accounted for. 
In fact, there is no need to extrapolate the pseudo-Hamiltonian
description to the fluctuation region around the critical temperature in view of the availability of the numerical answers for universal scaling functions describing all the weakly interacting U(1) systems at $|T-T_{\rm c}|/T_{\rm c} \ll 1$.  With these data, the two analytical descriptions---in the normal and superfluid phases---are readily 
connected across the fluctuation region.
By comparing our results to first-principles Mote Carlo data, we observe that the most vulnerable
region of parameters is $|T-T_{\rm c}|/T_{\rm c} \ll 1$, where the description is accurate only when the gas 
parameter $an^{1/3}$ is as small as $\sim 10^{-2}$.

\ack

We acknowledge support from NSF (Grant No. PHY-0653183), the Swiss National Science Fund, ITAMP, and the DARPA OLE program.


\end{document}